\documentstyle[epsfig,aps,prb,twocolumn,floats,eqsecnum]{revtex}

\begin{document}

\newcommand{\gsim}{
\,\raisebox{0.35ex}{$>$}
\hspace{-1.7ex}\raisebox{-0.65ex}{$\sim$}\,
}

\newcommand{\lsim}{
\,\raisebox{0.35ex}{$<$}
\hspace{-1.7ex}\raisebox{-0.65ex}{$\sim$}\,
}

\newcommand{\const}{ {\rm const} }
\newcommand{\arctanh}{ {\rm arctanh} }

\bibliographystyle{prsty}

\title{ \begin{flushleft}
{\small 
PHYSICAL REVIEW B 
\hfill \qquad\qquad
VOLUME  
{\normalsize 57}, 
NUMBER 
{\normalsize 21}
\hfill 
{\normalsize 1}
JUNE
{\normalsize 1998}-I, 
{\normalsize 13639$-$13654}
}\\
\end{flushleft}  
Quantum-classical transition of the escape rate of a uniaxial
spin system \\ in an arbitrarily directed field
}

\author{
D.~A. Garanin, \cite{e-gar}
X.~Mart\'\i nez~Hidalgo, \cite{e-mar}
and E.~M. Chudnovsky \cite{e-chu}
}

\address{
Department of Physics and Astronomy, City University of New York--Lehman College, \\
Bedford Park Boulevard West, Bronx, New York 10468-1589 \\
\smallskip
{\rm (Received 19 November 1997)}
\bigskip\\
\parbox{14.2cm}
{\rm
The escape rate $\Gamma$ of the large-spin model described by the Hamiltonian
${\cal H} = -DS_z^2 - H_zS_z - H_xS_x$ is investigated with the help
of the mapping onto a particle moving in a double-well potential $U(x)$.
The transition-state method yields $\Gamma$ in the moderate-damping
case as a Boltzmann average of the quantum transition probabilities.
We have shown that the transition from the classical to quantum regimes with 
lowering temperature is of the first order 
($d\Gamma/dT$ discontinuous at the transition temperature $T_0$)
for $h_x$ below the phase boundary line $h_x=h_{xc}(h_z)$, where 
$h_{x,z}\equiv H_{x,z}/(2SD)$, and of the second order above this 
line.
In the unbiased case ($H_z=0$) the result is $h_{xc}(0)=1/4$, i.e., one fourth of the metastability boundary 
$h_{xm}=1$, at which the barrier disappears.
In the strongly biased limit $\delta\equiv 1-h_z \ll 1$, one has 
$h_{xc} \cong(2/3)^{3/4}(\sqrt{3}-\sqrt{2})\delta^{3/2}\cong 0.2345 \delta^{3/2}$, which is about one half of the boundary value
$h_{xm} \cong (2\delta/3)^{3/2} \cong 0.5443 \delta^{3/2} $.
The latter case is relevant for experiments on small magnetic
particles, where the barrier should be lowered to achieve measurable
quantum escape rates.
\smallskip
\begin{flushleft}
\end{flushleft}
} 
} 
\maketitle

\section{Introduction}

The two fundamental mechanisms of the escape of a particle from a 
metastable potential well are quantum tunneling through the barrier 
and the classical thermal activation over the barrier.
The first mechanism is closely related to the tunneling level
splitting for a particle in a double-well potential, 
which was considered by Hund \cite{hun27} for the ammonia molecule.
Other early studies based on the WKB approximation 
\cite{wen26,kra26,bri26}
treated the ionization of atoms in electric fields,
\cite{opp28} cold emission of electrons from metal surfaces,
\cite{fownor28} and decay of nuclei. \cite{gam28}
Tunneling in spin systems was considered much later: Korenblit and
Shender \cite{korshe78} calculated the ground-state splitting in
the high-spin rare-earth compounds with the help of a high-order perturbation theory, Chudnovsky \cite{chu79} applied the instanton
technique for the Landau-Lifshitz equation to calculate the escape rates.
The current broad interest to the spin-tunneling problem was 
initiated, however, mainly by the application of the instanton method
by Enz and Schilling \cite{enzsch86} and Chudnovsky and Gunther, \cite{chugun88} and the spin-WKB formalism by van Hemmen and 
S\"ut\H o. \cite{lvhsuto86}        

Studying thermally activated escape of a classical particle
from the metastable minimum of a potential $U(x)$ goes back to 
Kramers, \cite{kra40} who solved the Fokker-Planck equation 
describing the diffusion of the particle over the barrier. 
For spin systems, the role of thermal agitation in overcoming
energy barriers (described, e.g., by the Stoner-Wohlfart model
\cite{stowoh4891}) was stressed by N\'eel. \cite{nee49}
Brown \cite{bro63} has derived the Fokker-Planck equation for
classical spin systems and calculated the escape rate in the uniaxial 
model.

An extensive reference to the thermal activation and tunneling of
particles can be found in Ref.\ \onlinecite{haetalbor90}, to magnetization tunneling in Ref.\ \onlinecite{stachubar92}, and to
the thermal activation in classical spin systems in 
Ref.\ \onlinecite{cofcalwal96book}.
Spin tunneling was recently observed in small magnetic particles such as ferritin \cite{awsetal92,gidetal95,tejetal97} 
and barium ferrite, \cite{weretal97q} and in high-spin
molecules, Mn$_{12}$Ac 
\cite{novses95,baretal95,frisartejzio96,heretal96,thoetal96} 
(see also 
Refs.\  \onlinecite{popular})
and Fe$_8$. \cite{sanetal97}

Considering escape at finite temperatures, the first idea is to 
sum the tunneling and thermoactivation escape rates as stemming from
independent channels: $\Gamma = \Gamma_q + \Gamma_{\rm th}$. 
Since the thermoactivation rate follows the very steep Arrhenius 
temperature dependence, 
$\Gamma_{\rm th} = \Gamma_0 \exp(-\Delta U/T)$,
the transition between quantum and classical regimes occurs at the 
temperature $T_0$ defined by $\Gamma_q = \Gamma_{\rm th}(T_0)$.
Writing $\Gamma_q = A\exp(-B)$, ignoring prefactors and equating
the exponents, one obtains the estimation
%
\begin{equation}\label{t0up0}
T_0^{(0)} = \Delta U/B,
\end{equation}
where the superscript at $T_0$ says that the ground-state tunneling
is considered. 
For $T > T_0^{(0)} $ one has practically 
$\Gamma \cong \Gamma_{\rm th}(T)$, 
whereas below the transition $\Gamma \cong \Gamma_q$ is independent of
temperature.
The transition between the two regimes occures on the temperature 
interval $\Delta T \sim T_0^{(0)}/B$.
Since $B\propto S$, this is much smaller than $T_0^{(0)}$ in the 
quasiclassical limit, $S\gg 1$.
The simple scenario above is the prototype for the so-called first-order quantum-classical transition of the escape rate, which is accompanied
by the discontinuity of $d\Gamma/dT$ at $T_0$.
  
It turns out, however, that for common metastable or double-well
potentials, such as cubic or quartic parabola, another scenario is realized.
Below the crossover temperature $T_0$ the particles cross the barrier
at the most favorable energy level $E(T)$ which goes down from 
the top of the barrier to the bottom of $U(x)$ with lowering temperature. 
Such a regime is called thermally assisted tunneling (TAT).
The transition from the classical regime to TAT is smooth, with no
discontinuity of $d\Gamma/dT$ at $T_0$, and the transition temperature
is given by \cite{gol59}
%
\begin{equation}\label{t02}
T_0^{(2)} = \tilde\omega_0/(2\pi),
\qquad \tilde\omega_0=\sqrt{|U''(x_{\rm sad})|/m}. 
\end{equation}
Here the superscript in $T_0$ denotes the second-order transition,
$x_{\rm sad}$ corresponds to the top of the barrier (the saddle point of the potential), and $\tilde\omega_0$ is the so-called instanton frequency.
In fact, the above formula is valid in the moderate-damping case, in
which the Kramers' result for the classical escape rate is independent
of the friction constant $\eta$ and coincides with that of the 
simple transition-state theory (TST).
In the strong- and weak-damping cases $\Gamma_0 \propto 1/\eta$ and
$\Gamma_0 \propto \eta$, respectively, \cite{kra40} and the formula
(\ref{t02}) should be modified 
(see, e.g., Ref.\ \onlinecite{haetalbor90}).
The consideration in the moderate-damping case is the most simple, 
and all the results can be obtained from the simple quantum TST
formula neglecting dissipation and giving the escape rate as a
Boltzmann average of quantum escape rates at different energies. 
\cite{bel5980,aff81}   
The quantum-classical transition of the escape rate including the
dissipation in the strong- to moderate-damping regimes was described
with the help of the Caldeira-Leggett formalism \cite{calleg83} in
Refs.\  \onlinecite{grawei84,graolswei85,larovc84,zwe85,rishaefre85}.
The results show that in the exponential approximation for the escape
rate only the second derivative $d^2\Gamma/dT^2$ is discontinuous at $T_0^{(2)}$.
More accurate calculations taking into account the prefactor 
\cite{aff81,graolswei85} show the smoothening of the transition in the
vicinity of $T_0^{(2)}$ due to quantum effects and the thermal 
distribution, so that all the derivatives of $\Gamma(T)$ behave 
continuously.

The terms first- and second-order quantum-classical transitions of
the escape rate used above are due to Larkin and Ovchinnikov. \cite{larovc83} 
Chudnovsky \cite{chu92} stressed the analogy with the phase
transitions and analyzed the general conditions for both
types of quantum-classical transitions.
He has shown that for the second-order 
transition the period of oscillations $\tau(E)$ in the inverted potential $-U(x)$ should monotonically increase with the amplitude
of oscillations, i.e., with the lowering energy $E$ from the top of the barrier.
If $\tau(E)$ is nonmonotonic, the first-order transition occurs.
Quite recently an effective free energy $F(E)$ for quantum-classical 
transitions of the escape rate of a spin system was written,
\cite{chugar97} the minimization of which
determines the escape rate in the exponential approximation:
$\Gamma \sim \exp(-F_{\rm min}/T)$.
The latter has the form $F = a\phi^2 + b\phi^4 + c\phi^6 +F_0$, 
just as in the Landau model of phase transitions. \cite{lan37} 
Here $a=0$ corresponds to the quantum-classical transition and 
$b=0$ to the boundary between first- and second-order transitions.

In a sense, second-order quantum-classical transitions of the escape rate are common, whereas the first-order ones are exotic and have to 
be specially looked for.
Nevertheless, a number of systems and processes showing first-order
transitions are already known, e.g., a SQUID with two Josephson junctions, \cite{morivlbla94} false vacuum decay in field theories, \cite{gar94,fer95,habmottin96,zimtchmue97} 
and depinning of a massive string from a linear defect.
\cite{skv97,gorbla97}
All these systems have more degrees of freedom than just a particle, 
thus the search for a physical system equivalent to a particle in a potential $U(x)$ leading to the first-order transition of the escape
rate seems to be quite actual.
Qualitatively it is clear how $U(x)$ should look: The top of the barrier should be rather flat, whereas the bottom should not.
In this case, as for the rectangular barrier, tunneling
just below the top of the barrier is unfavorable,
the TAT mechanism is suppressed, and the thermal activation competes
with the ground-state tunneling, leading to Eq.\  (\ref{t0up0}).  
Such a requirement is satisfied, e.g., for the pinning potential, \cite{skv97,gorbla97} which consists of 
periodically spaced narrow pits.
Here the qualitative results can be easily anticipated, at least for
particles moving in such a potential.
However, the exact form of the pinning potential is not known.

A rather simple and experimentally important system satisfying the
above requirement is the uniaxial spin model in a field described
by the Hamiltonian    
%
\begin{equation}\label{spinham}
{\cal H} = -DS_z^2 - H_zS_z - H_xS_x .
\end{equation}
This Hamiltonian can be mapped 
\cite{scharf74,zasulytsu83,schwrelvh87,zas90pla,zas90prb}
onto a particle moving in the potential $U(x)$ which has a 
double-well form in the region of field variables 
$\tilde h_{x,z}\equiv H_{x,z}/(2\tilde S D)$, 
$\tilde S \equiv S+1/2$ satisfying 
$\tilde h_x^{2/3}+ \tilde h_z^{2/3} \leq 1$, as the original spin
model (\ref{spinham}) in the classical limit.\cite{stowoh4891}    
The first-order escape-rate transition in the unbiased 
($H_z=0$) model (\ref{spinham}) for $H_x$ below some critical value
was found in Ref.\ \onlinecite{garchu97}.
In Ref.\ \onlinecite{chugar97} the exact value $\tilde h_{xc}=1/4$
was obtained with the help of the particle mapping.
One can get an idea of why the first-order transition should occur at
small $H_x$ from the following simple arguments.
Since tunneling in the model (\ref{spinham}) is caused entirely by the
transverse field $H_x$, it becomes very small for $\tilde h_x \ll 1$.
In this limit the barrier height $\Delta U$ remains finite, and the
form of the potential near the bottoms should also be preserved.
The only possibility for the vanishing tunneling rate is that the
barrier becomes very thick, with a very flat top 
(see, e.g.,  Fig.\  \ref{tatb_ux}).
The latter is just what is needed for the first-order 
quantum-classical escape rate transition.

The aim of this article is to generalize the approach for the biased
case $H_z\ne 0$ and to compute the entire phase diagram with the 
boundary line $\tilde h_{xc}(\tilde h_z)$ below which the transition
is first order.
We will use the simple damping-independent quantum TST formula as 
the starting point for calculations.
This requires justification for our spin system.
It is known that for the model without the transverse field the 
thermoactivation escape rate is proportional to the damping constant,
$\Gamma_0 \propto \eta$, for all values of $\eta$. \cite{bro63}
The models considering hopping over discrete levels for moderate 
values of $S$ yield the same result. \cite{vilharsesret94,gar97pre}
Such a situation can be thought of as the weak- and strong-damping regimes at the same time.
(This is not a contradiction, since according to the Landau-Lifshitz
equation, \cite{lanlif35} but not according to the Gilbert equation, \cite{gil55} larger values of $\eta$ always lead to a faster
relaxation.)
This situation is different from that of a particle, because, 
in terms of polar angles $\theta$ and $\varphi$,
in the axially symmetric case $H_x=0$
the spins cross the barrier not through the vicinity of a saddle
point but through the ridge $\theta=\theta^*$, 
where the energy of the spin has a maximum. 
If the transverse field is applied, the spins flow over the vicinity of a saddle point $\theta=\theta^*$, $\varphi=0$.
This brings the system closer to the usual situation with particles,
and the moderate-damping regime with the damping-independent 
$\Gamma_0$ appears. 
\cite{smiroz76,bro79,kligun90,geocofmul97,cofetal97}
The crossover from the strong- to moderate-damping regimes was
confirmed recently by a numerical solution of the Fokker-Planck equation for classical spins in the oblique field in 
Ref.\ \onlinecite{cofetal97}.
The boundaries of the moderate-damping regime for the spin system
depend, in addition to $\eta$, on $H_x$ and $H_z$ and they are 
not yet well established.
Accordingly, an accurate description of spin tunneling with 
dissipation is an open problem.
For this reason we restrict ourselves in this work to the simple
damping-independent quantum TST approach.
     
The structure of the main part of this article is the following.
In Sec. \ref{fundamentals} the fundamentals concerning the particle 
mapping, WKB approximation and the quantum TST are reviewed.
In Sec. \ref{unbiased} the quantum-classical transition of the escape
rate in the unbiased case is studied, and the escape rate is 
calculated in the whole temperature range including the prefactor.
In Sec. \ref{biased} the results are generalized for the biased case.
The possibilities of the experimental observation of different types
of the escape-rate transition are discussed in 
Sec. \ref{discussion}.

\section{Particle mapping and escape rate}
\label{fundamentals}

\subsection{Particle mapping}
\label{mapping}

The spin problem with the Hamiltonian (\ref{spinham}) can be mapped
\cite{scharf74,zasulytsu83,schwrelvh87,zas90pla,zas90prb}
onto a particle problem
%
\begin{equation}\label{partpro}
{\cal H} = \frac{p^2}{2m} + U(x), \qquad p=-i\frac{d}{dx},
\qquad m=\frac{1}{2D} . 
\end{equation}
The mapping makes a correspondence between the spin wave function
$\Psi_S = \sum_{m=-S}^S C_m |m\rangle$, where $|m\rangle$ are the 
eigenstates of $S_z$, and the coordinate wave function
%
\begin{equation}\label{psix}
\Psi(x) = e^{-f(x)}\sum_{m=-S}^S \frac{ C_m \exp(mx) }
{ \sqrt{(S-m)!(S+m)!} } ,
\end{equation}
where
%
\begin{equation}\label{fxmap}
f(x) = \tilde S[\tilde h_x \cosh(x) - \tilde h_z x]
\end{equation}
and 
%
\begin{equation}\label{tildef}
\tilde S \equiv S+1/2, 
\qquad \tilde h_{x,z} \equiv H_{x,z}/(2\tilde S D).
\end{equation}
It can be shown that if the coefficients $C_m$ satisfy the 
Schr\"odinger equation for the spin problem (\ref{spinham}), the 
coordinate wave function (\ref{psix}) satisfies the Schr\"odinger
equation for the particle problem (\ref{partpro}), in the stationary
case with the same energy levels $E_n$, $n=0,1,\ldots,2S$.
The potential $U(x)$ in Eq. (\ref{partpro}) is given by
%
\begin{eqnarray}\label{ux}
&&
U(x) = \tilde S^2 D u(x),
\nonumber\\
&&
u(x) = [\tilde h_x \sinh(x) - \tilde h_z]^2 - 2\tilde h_x\cosh(x).
\end{eqnarray}
It has a double-well form for $H_x$ and $H_z$ inside the region 
closed by the metastability boundary curve
\cite{stowoh4891}
%
\begin{equation}\label{metbou}
\tilde h_{xm}^{2/3} + \tilde h_{zm}^{2/3} = 1 .
\end{equation}
This simple formula derived 50 years ago has been only recently 
tested in experiment on individual single-domain particles. \cite{weretal97} 

\begin{figure}[t]
\unitlength1cm
\begin{picture}(11,7)
\centerline{\epsfig{file=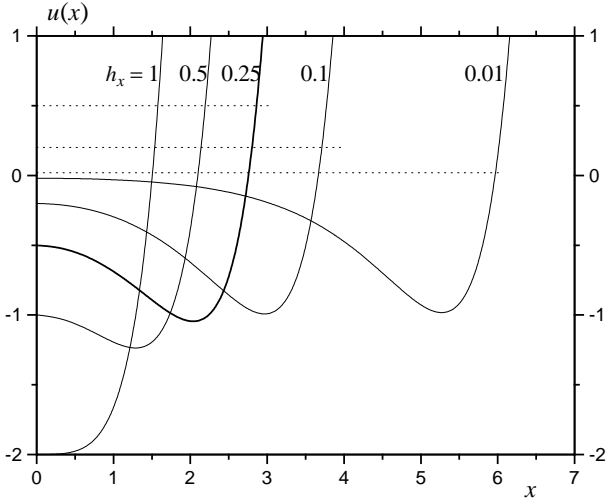,angle=-90,width=12cm}}
\end{picture}
\caption{ \label{tatb_ux} 
Reduced effective potential for the spin system, 
Eq.\  (\protect\ref{ux}), in the unbiased case.
The boundaries between the spin states and unphysical states are
indicated by horizontal dotted lines.
}
\end{figure}

Finding the extrema of $U(x)$ requires the solution of the fourth-order
algebraic equation for $y=\exp(x)$ and it can be better done 
numerically.
In the unbiased case, $H_z=0$, the top of the barrier is at $x=0$
and it corresponds to the saddle point of the classical spin 
Hamiltonian (\ref{spinham}).
The mininum of $U(x)$ is attained at $\tilde h_x\cosh(x_{\rm min})=1$.
One has 
%
\begin{equation}\label{uextr}
U_{\rm sad} = -2\tilde S^2 D\tilde h_x,
\qquad U_{\rm min} = -\tilde S^2 D(1+\tilde h_x^2) ,
\end{equation}
which yields the barrier height
%
\begin{equation}\label{deltau}
\Delta U \equiv U_{\rm sad} - U_{\rm min} = 
\tilde S^2 D(1-\tilde h_x)^2,
\end{equation}
as for the original spin problem.
The frequency of small oscillations near the bottom of $U(x)$
(the attempt frequency) 
%
\begin{equation}\label{om0}
\omega_0=\sqrt{U''(x_{\rm min})/m} = 2\tilde SD\sqrt{1-\tilde h_x^2} 
\end{equation}
coincides with the ferromagnetic resonance frequency in the classical
limit $S\to\infty$.
The instanton frequency $\tilde \omega_0$ of Eq.\  (\ref{t02}) reads
%
\begin{equation}\label{tilom0}
\tilde \omega_0 = 2\tilde SD\sqrt{\tilde h_x(1-\tilde h_x)} . 
\end{equation}
It becomes much smaller than $\omega_0$ for $\tilde h_x\ll 1$, which
signifies the flat top of the barrier in this limit. 
The expansion of the potential near the top of the barrier in the
unbiased case has the form 
%
\begin{eqnarray}\label{uxsmall}
&&
u(x) \cong -2\tilde h_x - \tilde h_x(1-\tilde h_x)x^2
+\frac{\tilde h_x}{3}\left(\tilde h_x-\frac14\right)x^4
\nonumber\\
&&
\qquad
{}+ \frac{2\tilde h_x}{45}\left(\tilde h_x-\frac{1}{24}\right)x^6
 + \ldots ,
\end{eqnarray}
where the change of the sign of the coefficient in  
the $x^4$ term is responsible for the first-order escape-rate
transition at $\tilde h_x=1/4$. \cite{chugar97}
The behavior of $u(x)$ for different values of the transverse field 
is represented in Fig.\  \ref{tatb_ux}.
\par
The particle wave functions (\ref{psix}) imaging those of the spin
system are proportional to polynomials of power $2S$ in $y=\exp(x)$,
thus they can have up to $2S$ zeros and they describe $2S+1$ spin 
states.
The corresponding solutions of the particle's Schr\"odinger equation 
for the potential (\ref{ux}) in the unbiased case were studied by
Razavy, \cite{raz80} without a reference to the spin problem.
He has found explicit analytical solutions corresponding to
$S=1/2,1$, and 3/2, as well as formally to $S=0$.
The particle problem (\ref{partpro}) possesses, however, an infinite
number of states with the energy eigenvalues above those of the 
spin problem. 
These additional unphysical states could, in principle, mix together with the true spin states and affect the results.
This does not happen if the unphysical states are much less thermally
populated than the spin states near the top of the barrier,
$E=U_{\rm sad}=-2S^2 D h_x$ [here $h_x$ is defined by
Eq. (\ref{tildef}) without the tilde].
The criterium for neglecting the unphysical states can be formulated,
if one notices that the boundary between the two groups of states corresponds, in the quasiclassical limit, to the maximal energy of the spin model
(\ref{spinham}),
%
\begin{equation}\label{usup}
U_{\rm max}= 2S^2 D h_x.
\end{equation}
Thus, the unphysical states do not affect the results and the analogy
with the particle is complete if
%
\begin{equation}\label{mapcond}
T \ll U_{\rm max} - U_{\rm sad} = 4S^2 D h_x . 
\end{equation}
This criterium can be violated, if the field $h_x$ is small and 
temperature is not low enough.
However, we are mainly interested here in the temperature range about
the quantum-classical transition temperature $T_0\sim SD$.
For such temperatures the complete analogy with the particle is 
justified by the large spin value $S$.

\subsection{Escape rate and level splitting}
\label{rate}

The simple quantum transition-state theory postulates the escape rate
in the quasiclassical case and in the temperature range 
$T\ll \Delta U$ in the plausible form
%
\begin{eqnarray}\label{tst}
&&
\Gamma = \frac{1}{Z_0} \sum_n \Gamma_n
\exp\left(-\frac{ E_n - U_{\rm min} }{ T }\right) ,
\nonumber\\
&&
\Gamma_n = \frac{ \omega(E_n) }{ 2\pi } W(E_n) , 
\qquad Z_0 \cong \frac{ 1 }{ 2\sinh[\omega_0/(2T)]} ,
\end{eqnarray}
where $\omega(E_n)$ is the frequency of oscillations at the energy level $E_n$, $W(E_n)$ are quantum transition probabilities, and
$Z_0$ is the partition function in the well calculated for 
$T\ll \Delta U$ over the low-lying oscillatorlike states with
$E_n = (n+1/2)\omega_0 + U_{\rm min}$.
Since the transition probability $W(E_n)$ usually increases rapidly
with energy $E$, the sum in Eq.\  (\ref{tst}) extends for not too low
temperatures over many levels and can be replaced by the integral
according to $\sum_n \ldots \Rightarrow \int dE \rho(E) \ldots$
with the density of states $\rho(E)=1/\omega(E)$. \cite{lanlif3}
This leads to 
%
\begin{equation}\label{tst1}
\Gamma = \frac{1}{2\pi Z_0} \int_{U_{\rm min}}^{U_{\rm max}}\! \!
dE\, W(E)
\exp\left(-\frac{ E-U_{\rm min} }{ T }\right) .
\end{equation}
For particles ${U_{\rm max}} = \infty$, and  
the usual expression \cite{bel5980,aff81} is recovered.
The barrier transparency $W(E)$ is determined in the WKB approximation by the imaginary-time action 
(see, e.g., Ref.\ \onlinecite{lanlif3})
%
\begin{equation}\label{action}
S(E) = 2\sqrt{2m}\int_{x_1(E)}^{x_2(E)} dx\; \sqrt{U(x)-E},
\end{equation}
where $x_{1,2}(E)$ are the turning points for the particle
oscillating in the inverted potential $-U(x)$.
The factor 2 in Eq.\  (\ref{action}) says that the integral is taken
over the whole period of oscillations, i.e., the particle crosses
the barrier twice.
The WKB approximation fails for energies near the top of the barrier,
but the result for $W(E)$ can be improved and extended for the energies above the barrier, if the barrier is
parabolic near the top. \cite{kem35,lanlif3,aff81} 
The result can be conveniently written as 
%
\begin{equation}\label{we}
W(E) = \frac{1}{1 + \exp[S(E)]} ,
\end{equation}
where $S(E)$ goes linearly through zero for $E$ crossing the barrier
top level and it should be analytically continued into the region
$E>U_{\rm sad}$.
In the latter case formula (\ref{we}) describes quantum 
reflections for a particle going over the barrier, with $W(E)$
slightly lower than 1. 
In the classical limit $W(E)\Rightarrow\theta(E-U_{\rm sad})$, where 
$\theta(x)$ is the step function, whereas the partition function $Z_0$
given by Eq.\  (\ref{tst}) simplifies to $T/\omega_0$.
Then integration in Eq.\  (\ref{tst1}) yields the formula
%
\begin{eqnarray}\label{tstcl}
&&
\Gamma = \Gamma_0 \exp(-\Delta U/T) ,
\nonumber\\
&& 
\Gamma_0 = \frac{ \omega_0 }{2\pi } 
\left[ 1 - \exp\left(-\frac{ U_{\rm max} - U_{\rm sad} }{ T }\right)
\right] .
\end{eqnarray}
If the condition (\ref{mapcond}) is satisfied, the second
exponential in Eq.\  (\ref{tstcl}) can be neglected, and the well-known
moderate-damping result for particles \cite{kra40} is recovered.
In the opposite case the prefactor $\Gamma_0$ in 
Eq.\  (\ref{tstcl}) becomes proportional to $h_x$ and vanishes in the 
limit $h_x\to 0$.
In fact, in this case dissipation should be taken into account,
and the well-known result is $\Gamma_0 \propto \eta$ for all values of 
the damping constant $\eta$. \cite{bro63}
The meaning of this result is that in the axially symmetric case
precession of a spin around the anisotropy axis does not bring it closer to the barrier, and the role of $\omega(E)$ as the attempt 
frequency is lost.
In this situation spin can overcome the barrier only via the diffusion
in the energy space.

The level splitting $\Delta E_n$ is related to the barrier 
transparency $W(E_n)$, and in the WKB approximation it is given by
\cite{lanlif3}
%
\begin{equation}\label{splt}
\Delta E_n =
\frac{\omega(E_n)}{\pi}\exp\left[-\frac{S(E_n)}{2}\right] .
\end{equation}
For the lowest energy levels this WKB result becomes invalid.
A more accurate consideration of Weiss and Haeffner \cite{weihae83}
using the functional-integral technique, as well as that of Shepard
\cite{she83} improving the usual WKB \cite{wen26,kra26,bri26} scheme
shows that for the potentials 
parabolic near the bottom the result above should be multiplied by
%
\begin{equation}\label{weicor}
\sqrt{\frac{\pi}{e}} \frac{ (2n+1)^{n+1/2} }{ 2^n e^n n! } .
\end{equation}
This factor approaches 1 as $1+1/(24n)$ with increasing $n$ and it is, in fact, very close to 1 for all $n$: 1.0750 for $n=0$, 1.0275 for $n=1$, 1.01666 for $n=2$, 1.01192 for $n=3$, etc.
Equating the expression in Eq.\  (\ref{weicor}) to 1 defines an approximation for $n!$ which is more accurate than the leading
term of the Stirling formula. 
The correction factor (\ref{weicor}), however, 
does not change much in front of the exponentially small action term
in Eq.\  (\ref{splt}).

The formula (\ref{splt}) has the meaning of the level splitting only
if the levels in the two wells are degenerate.
More generally, Eq.\  (\ref{splt}) yields the transition amplitude
between the levels on both sides of the barrier,
and the expression for the escape rate should be sensitive to the
resonance condition for the levels in the two wells.
Therefore, the quantity $\Gamma_n$ 
in Eq.\  (\ref{tst}) should be replaced by \cite{garg95diss,garchu97}
%
\begin{equation}\label{lorenz}
\Gamma_n =\frac{(\Delta E_n)^2}{2} \sum_{n'} 
\frac{ \Gamma_{nn'} }{ (E_n-E_{n'})^2 + \Gamma_{nn'}^2 } ,
\end{equation}
where $\Delta E_n$ is given by Eq.\  (\ref{splt}), $n'$ are the levels
in the other well and $\Gamma_{nn'}$ is the sum of the linewidths of
the $n$th and $n'$th levels.
If $\Gamma_{nn'}$ substantially exceeds the level spacing $\omega_{n'}\equiv E_{n'+1} -E_{n'}$, the sum
in Eq.\  (\ref{lorenz}) can be approximated by the integral and one obtains
%
\begin{equation}\label{gamspl}
\Gamma_n = \pi (\Delta E_n)^2/[2\omega(E_n)].
\end{equation}
The latter is by virtue of Eq.\  (\ref{splt}) equivalent to $\Gamma_n$
in Eq.\  (\ref{tst}), and the reasoning above can be considered as a 
derivation of it.
However, such a situation implies that the system is so strongly
damped that the free precession of the spin decays long before the
period of the precession is completed.
This is hardly the case for magnetic systems, which is known both from 
experiments and from theoretical estimations. 
\cite{garkim8991}
Indeed, resonant tunneling was observed in experiments
on high-spin magnetic molecules, \cite{frisartejzio96} as well as on 
{\em underdamped} Josephson junctions.  
\cite{roshanluk95,silpalrugrus97}
In this paper we will, however, ignore resonance effects in
spin tunneling, which have been considered in more detail in 
Ref.\ \onlinecite{garchu97}.  

Formula (\ref{tst}) describes the escape from the metastable (left)
to the stable (right) well.
In the unbiased or weakly biased cases the rate of exchange between
the two wells, which is the observable quantity, is the sum of the
escape rates from both wells into the opposite one
(see, e.g., Ref.\ \onlinecite{aha69}).
Thus, in the unbiased case all the results for $\Gamma$ below should be multiplied by 2, which will be kept in mind but not done 
explicitly.
For low temperatures it is sufficient to introduce a small bias field
to suppress the backflow from the stable to the metastable well and to
make the consideration neglecting this process absolutely correct.
The condition for this is $\xi\equiv SH_z/T \gsim 1$.

\section{Quantum-classical transition in the unbiased case}
\label{unbiased}

\subsection{Level splitting and transition probability}
\label{splitting}

In the unbiased case, $H_z=0$, the imaginary-time action $S(E)$
of Eq.\  (\ref{action}) can be reduced to elliptic integrals, e.g., 
by the substitution $y=\cosh(x)$.
The appropriate formula is available in Ref.\ \onlinecite{byrfri79},
and the result is
%
\begin{equation}\label{seres}
S(E) = 4\tilde S [ (1-\tilde h_x) f_+ ]^{1/2}
I(\alpha^2,k) ,
\end{equation}
where
%
\begin{equation}\label{fpmdef}
f_\pm \equiv P + \tilde h_x(1\pm\sqrt{1-P})^2 
\end{equation}
and
%
\begin{equation}\label{pdef}
P\equiv  (U_{\rm sad} - E )/( U_{\rm sad} - U_{\rm min} ) 
\end{equation}
[see Eq.\  (\ref{uextr})]
is the dimensionless energy variable taking the values 0 at the top 
of the barrier and 1 at the bottom of the potential.
The quantity $I(\alpha^2,k)$ above is given by
%
\begin{eqnarray}\label{ink}
&&
I = (1+\alpha^2)\bbox{K} - \bbox{E} + 
(\alpha^2-k^2/\alpha^2)[\Pi(\alpha^2,k)-\bbox{K}],
\nonumber\\
&&
\alpha^2 = (1-\tilde h_x)P/f_+, \qquad k^2 = f_-/f_+ ,
\end{eqnarray}
where $\bbox{K}$ and $\bbox{E}$ are complete elliptic integrals of
the first and second kind of modulus $k$ and $\Pi(\alpha^2,k)$ 
is the complete elliptic integral of the third kind.
The latter can be expressed through the incomplete elliptic integrals of the first and second kinds, in our case of $\alpha^2<k^2$, as 
\cite{byrfri79}  
%
\begin{eqnarray}\label{p12}
&&
\Pi(\alpha^2,k) - \bbox{K} = 
\frac{ \alpha [ \bbox{K} E(\beta,k) - \bbox{E} F(\beta,k) ] }
{ \sqrt{ (1-\alpha^2)(k^2 - \alpha^2) } } ,
\nonumber\\
&&
\beta = \arcsin(\alpha/k). 
\end{eqnarray}

The general expression for $S(E)$ simplifies in five limiting cases.
The first case is $1-\tilde h_x \ll 1$, in which $U(x)$ reduces
to the quartic parabola [see Eq.\  (\ref{uxsmall}) 
and Fig.\  \ref{tatb_ux}].
Here the parameters $\alpha$ and $k$ of Eq.\  (\ref{ink})
simplify to
%
\begin{equation}\label{alpkqp}
\alpha^2 \cong \frac{ (1-\tilde h_x)P }{ 2(1+\sqrt{1-P}) },
\qquad k^2 \cong \frac{ 1-\sqrt{1-P} }{ 1+\sqrt{1-P} }.
\end{equation}
Expanding Eqs.\  (\ref{ink}) and (\ref{p12}) in powers of 
$\alpha^2\ll 1$ one obtains  \cite{she83}
%
\begin{eqnarray}\label{seqp}
&&
S(E) \cong (2\tilde S/3) [ 2(1-\tilde h_x)(1+\sqrt{1-P}) ]^{3/2}
\nonumber\\
&&
\qquad\qquad
{}\times [(1+k^2) \bbox{E} - (1-k^2) \bbox{K} ] .
\end{eqnarray}

For small transverse fields and the energies near the top of the 
barrier, $\tilde h_x \sim P \ll 1$, one can use
$f_+\cong P+4\tilde h_x$ and $f_-\cong P$, which results in
$k^2 \cong \alpha^2 \cong P/(P+4\tilde h_x)$ and 
$k^2/\alpha^2 - \alpha^2 \cong 1-k^2$. 
Now with the help of $\Pi(k^2,k)=\bbox{E}/(1-k^2)$ one arrives at 
the result
\cite{lvhsuto86}
%
\begin{equation}\label{sehsmall}
S(E) \cong 8\tilde S [P+4\tilde h_x]^{1/2} (\bbox{K} - \bbox{E}).
\end{equation}
Here for $4\tilde h_x \ll P \ll 1$ the modulus $k$ of the elliptic integrals is close to 1.
In this case using $\bbox{K}\cong \ln(4/\sqrt{1-k^2})$ and 
$\bbox{E}\cong 1$, as well as $P\cong |E|/(\tilde S^2D)$, one can 
simplify Eq.\  (\ref{sehsmall}) to \cite{lvhsuto86}
%
\begin{equation}\label{sehsmallp}
S(E) \cong 4 \sqrt{ \frac{|E|}{D} }
\ln\left(\frac{8}{e^2}\frac{E}{U_{\rm sad}}\right), 
\end{equation}
where $U_{\rm sad}$ is given by Eq.\  (\ref{uextr}).
Now one can rewrite the level splitting (\ref{splt}) introducing the
unperturbed energy levels $E_m=-Dm^2$, $m=-S, -S+1,\ldots,S$ and
ignoring the prefactor in the form
%
\begin{equation}\label{spltm}
\Delta E_m \sim (m_b/m)^{4m}, 
\qquad m_b^2 = 2\tilde S^2\tilde h_x(e^2/8) ,
\end{equation}
which is valid for $1 \ll m \ll S$.
The latter is the quasiclassical limit of the perturbative 
in $\tilde h_x$ result of 
Refs.\ \onlinecite{gar91jpa,harbou95}, and \onlinecite{garchu97}.

One can try to continue the perturbative formulas (\ref{sehsmallp})
and (\ref{spltm}) to the very top of the barrier, although the 
perturbation theory breaks down there.
It can be seen that, in contrast to $S(E)$ of Eq.\  (\ref{sehsmall}) turning to
zero at $E/U_{\rm sad}=1$, Eq.\  (\ref{sehsmallp}) turns to
zero at $E/U_{\rm sad}=e^2/8=0.92$.
Accordingly, in Eq.\  (\ref{spltm}) $\Delta E_m \sim 1$ at $m=m_b$ and
not at $m^2=2\tilde S^2\tilde h_x$, which corresponds to  
$E_m=-Dm^2=U_{\rm sad}$.
This can be interpreted as a shift of the barrier height by the
artifact $e^2/8$ in the perturbative formalism.
This factor is close to unity and is not very important.
However, the inaccuracy of the perturbative method manifests itself
much stronger in the quantum-classical transition of the escape rate.
The perturbative result of Ref.\ \onlinecite{garchu97} for the boundary
between the first- and second-order transitions is 
$\tilde h_{xc}\approx 0.13$, which is about a half of the
actual boundary value $\tilde h_{xc}=1/4$. \cite{chugar97}

Another limiting case of Eq.\  (\ref{seres}) is the case of a small 
transverse field and energy not too close to the top of the barrier,
$4\tilde h_x \ll P \sim 1$.
This is a purely perturbative case, and the level splitting for 
the energy levels labeled by $m$ is given by the formula
\cite{gar91jpa}
%
\begin{equation}\label{spltmgen}
\Delta E_m \cong \frac{ 2D }{ [(2|m|-1)!] }
\frac{ (S+|m|)! }{ (S-|m|)! } 
\left( \frac{ H_x }{ 2D } \right)^{2|m|} .
\end{equation}
According to Eq.\  (\ref{splt}) the imaginary-time action $S(E)$ 
can be obtained from the above formula as 
$S(E)= -2\ln(\pi\Delta E_m/|\omega_{m+1,m}|)$, 
where the energy variable $P=m^2/S^2$ should be introduced.
The WKB approximation for $S(E)$ in this case, which interpolates
between Eqs.\  (\ref{sehsmallp}) and (\ref{suminpert}), was calculated
in Ref.\ \onlinecite{schwrelvh87} with the help of the particle mapping
and it can be found there.

The other two cases in which $S(E)$ simplifies are those 
corresponding to the energy near the top of the barrier or the bottom of the well.
For the parametrization it is convenient to introduce the dimensionless quantities
%
\begin{equation}\label{vtildef}
\tilde v\equiv \Delta U/\tilde\omega_0 = 
\tilde S (1-\tilde h_x)^{3/2}/(2\tilde h_x^{1/2}) 
\end{equation}
and
%
\begin{equation}\label{vdef}
v\equiv \Delta U/\omega_0 = 
\tilde S (1-\tilde h_x)^{3/2}/[2(1+\tilde h_x)^{1/2}]
\end{equation}
[see Eqs.\  (\ref{deltau}), (\ref{om0}), and (\ref{tilom0})].
The quantity $v$, in particular, is a rough estimate of the number of
levels in the well, based on the assumption that the levels remain
equidistant up to the top of the barrier. 
It thus measures quantum effects in the system.
One can see that for $S\gg 1$ the system can be made more quantum by
applying a field close to the metastability boundary,
$1-\tilde h_x \ll 1$.
The condition $v\gg 1$, which also entails $\tilde v\gg 1$, should be,
however, satisfied for the applicability of the quasiclassical
method.
The quantity $\tilde v$ is related, as we will see immediately, to the
quantum penetrability of the barrier near its top and it also measures quantum effects.
For conventinal potentials, such as cubic or quartic parabola, 
$v$ and $\tilde v$ can differ only by a numerical factor, and it is 
sufficient to introduce one of them 
(see, e.g., Ref.\ \onlinecite{graolswei85}).   
This is not the case for our spin model for small transverse fields,
$\tilde h_x\ll 1$.

\begin{figure}[t]
\unitlength1cm
\begin{picture}(11,7)
\centerline{\epsfig{file=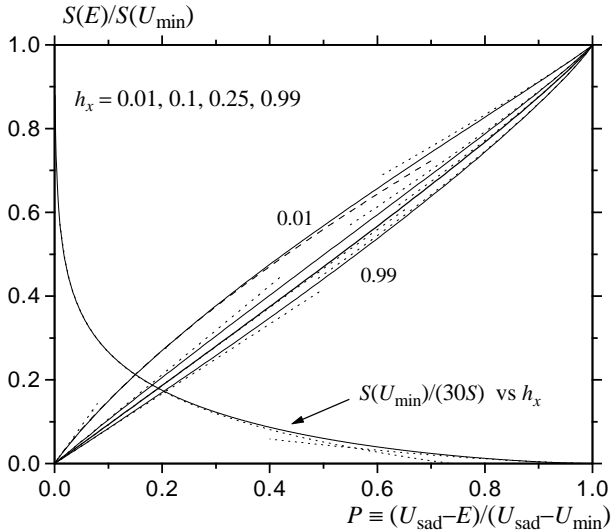,angle=-90,width=12cm}}
\end{picture}
\caption{ \label{tatb_se} 
Imaginary-time action $S(E)$ of Eq.\  (\protect\ref{seres}) 
for $\tilde h_z=0$ and different values of 
$\tilde h_x$ and the bottom-energy
action $S(U_{\rm min})$ vs $\tilde h_x$. 
The asymptotes (\protect\ref{setop}) and (\protect\ref{sebot}) for
$S(E)$, as well as 
$S^{\rm pert}(U_{\rm min})=4\tilde S \ln[2/(e\tilde h_x)]$  and 
Eq.\  (\protect\ref{sumineps}) for 
$S(U_{\rm min})$ vs $\tilde h$ are shown by the dotted lines.
The result of Eq.\  (\protect\ref{sehsmall}) is plotted for 
$\tilde h_x = 0.01$ with the dashed line.
}
\end{figure}

Near the top of the barrier, $P\ll 1$, the result for $S(E)$ following 
from Eq.\  (\ref{seres}) or directly from the integral expression
(\ref{action}) reads
%
\begin{equation}\label{setop}
S(E) \cong 2\pi\tilde v [P + bP^2 + cP^3 + O(P^4)]
\end{equation}
with
%
\begin{equation}\label{bcdef}
b = \frac{1}{8} \left(1-\frac{1}{4\tilde h_x}\right),
\qquad c =  \frac{3}{64} 
\left(1-\frac{1}{3\tilde h_x}+\frac{1}{16\tilde h_x^2}\right) .
\end{equation}
One can see that $b$ changes sign at $\tilde h_x=1/4$, whereas
$c>0$ for all $\tilde h_x$.
For small $\tilde h_x$ the coefficients $b$, $c$, etc., become large,
which means that Eq.\  (\ref{setop}) is only applicable for 
$P\ll \tilde h_x$.
For $\tilde h_x \sim P\ll 1$ Eq.\  (\ref{sehsmall}) can be used. 
The latter gives, however, accurate results only for 
$\tilde h_x\lsim 0.02$ (see Fig.\  \ref{tatb_se}).  
 
Near the bottom of the potential one obtains
%
\begin{equation}\label{sebot}
S(E) \cong S(U_{\rm min}) - 
\frac{ 2(E - U_{\rm min}) }{ \omega_0 }
\ln\left(\frac{ eq \omega_0 }{ E - U_{\rm min} }\right)
\end{equation}
with
%
\begin{equation}\label{sumin}
S(U_{\rm min})= 4\tilde S 
\left[ \ln \left( 
\frac{ 1 + \sqrt{1-\tilde h_x^2} }{ \tilde h_x }
\right) - \sqrt{1-\tilde h_x^2} \right] 
\end{equation}
and 
%
\begin{equation}\label{qdef}
q = 8 \tilde S (1-\tilde h_x^2)^{3/2}/\tilde h_x^2 .
\end{equation}
For $\tilde h_x \ll 1$ the bottom-level action $S(U_{\rm min})$ simplifies to 
%
\begin{equation}\label{suminpert}
S(U_{\rm min})\cong 4\tilde S 
\{ \ln[2/(e\tilde h_x)] + \tilde h_x^2/4 \}, 
\end{equation}
where the first term is the perturbative result, 
$S^{\rm pert}(U_{\rm min})$ 
(see Refs.\  \onlinecite{korshe78,gar91jpa}, and \onlinecite{harbou95}).
Since the correction term in Eq.\  (\ref{suminpert}) is quadratic
in $\tilde h_x$ with a small coefficient, $S^{\rm pert}(U_{\rm min})$ 
is a good approximation to $S(U_{\rm min})$ in the whole region
of the first-order escape-rate transition, $h_x<1/4$. 
In the other limiting case one has
%
\begin{equation}\label{sumineps}
S(U_{\rm min})\cong (4\tilde S/3) (2\epsilon)^{3/2},
\qquad \epsilon \equiv 1 - \tilde h_x\ll 1 
\end{equation}
(see, e.g., Ref.\ \onlinecite{chugun88}).
The dependences $S(E)$ for different values of $\tilde h_x$,
as well as $S(U_{\rm min})$ vs $\tilde h_x$ are
shown in Fig.\  \ref{tatb_se}.
\begin{figure}[t]
\unitlength1cm
\begin{picture}(11,7)
\centerline{\epsfig{file=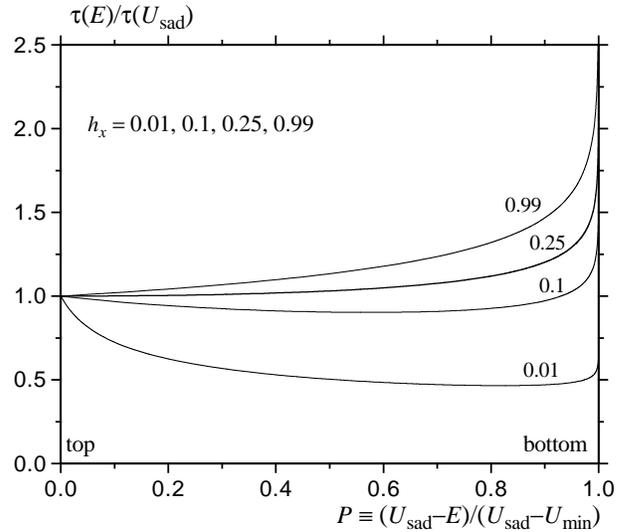,angle=-90,width=12cm}}
\end{picture}
\caption{ \label{tatb_tau} 
Instanton period $\tau(E)$ of Eq.\  (\protect\ref{taures}) 
for $\tilde h_z=0$ and different values of $\tilde h_x$. 
}
\end{figure}
\par

The instanton period
%
\begin{equation}\label{taudef}
\tau(E) = -\frac{dS(E)}{dE} = 
\sqrt{2m}\int_{x_1(E)}^{x_2(E)}\; \frac{dx}{\sqrt{U(x)-E}}
\end{equation}
can be easier calculated directly than by the differentiation of 
Eq.\  (\ref{seres}).
The result has the form
%
\begin{equation}\label{taures}
\tau(E) = [(1-\tilde h_x) f_+ ]^{-1/2} 4\bbox{K}/(\tilde SD) ,
\end{equation}
where both $f_+$ and $\bbox{K}$ have the same meaning as above 
[see Eqs.\  (\ref{fpmdef}) and (\ref{ink})]. 
Near the top of the barrier, $P\to 0$, one has $f_+\to 4\tilde h_x$,
$k\to 0$, $\bbox{K}\to \pi/2$, and Eq.\  (\ref{taures}) yields the previously known result $\tau=2\pi/\tilde\omega_0$ with $\tilde\omega_0$ given by Eq.\  (\ref{tilom0}).
Near the bottom, $P\to 1$, one has $k\to 1$, and $\tau$ 
logarithmically diverges. 
The dependences $\tau(E)$ for different values of $\tilde h_x$ are
shown in Fig.\  \ref{tatb_tau}.

One can see that for $\tilde h_x\geq 1/4$ the period $\tau(E)$
monotonically
increases with the amplitude of the oscillations in the inverted
potential $-U(x)$, i.e., with the increase of $P$ of Eq.\  (\ref{pdef}).
In this case the quantum-classical transition is second order.
\cite{chu92}
For $\tilde h_x < 1/4$ the dependence $\tau(E)$ is nonmonotonic,
and the transition is first order.
Such a behavior of $\tau(E)$ can be easily explained qualitatively.
For $h_x>1/4$ the fourth-order term in Eq.\  (\ref{uxsmall}) is
positive, i.e., $U(x)$ is of the form $-x^2+x^4$.
The inverted potential $-U(x)$ is hence of the type $x^2-x^4$,
which results in the increase of $\tau$ with the oscillation
amplitude (i.e., with lowering the energy $E$) and to the second-order 
transition.
At $h_x<1/4$ the anharmonicity of $-U(x)$ has the
opposite sign, $-U(x) \sim  x^2+x^4$, which leads to the
decrease of $\tau$ when lowering $E$ for energies below the
top of the barrier.
However, with further lowering of $E$ the period $\tau$ begins to
increase and diverges logarithmically for $E$ approaching the
bottom of the potential.

\begin{figure}[t]
\unitlength1cm
\begin{picture}(11,7)
\centerline{\epsfig{file=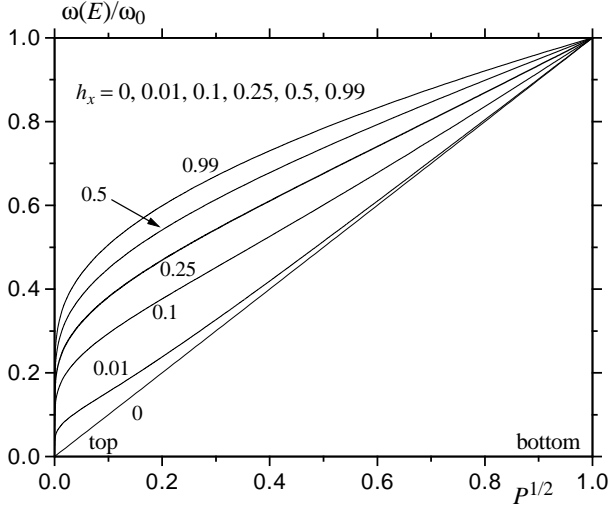,angle=-90,width=12cm}}
\end{picture}
\caption{ \label{tatb_ome} 
Frequency of the oscillations in the potential wells $\omega(E)$ of
Eq.\  (\protect\ref{omeres}) 
for $h_z=0$ and different values of $h_x$. 
}
\end{figure}

The period of the real-time oscillations in the potential minima
can be calculated with the use of the formula differing from
Eq.\  (\ref{taudef}) by changing sign under the square root.
The corresponding energy-dependent frequency has the form
%
\begin{equation}\label{omeres}
\omega(E) = 2\tilde SD [ (1-\tilde h_x) f_+ ]^{1/2} \pi/(2\bbox{K}) ,
\end{equation}
where
%
\begin{equation}\label{komdef}
\bbox{K} \equiv \bbox{K}(r), 
\qquad r^2 = 4\tilde h_x \sqrt{1-P}/f_+ ,
\end{equation}
and $f_+$ is given by Eq.\  (\ref{fpmdef}).
Near the bottom, $P\to 1$, one has $f_+\to 1+\tilde h_x$, $r\to 0$,
$\bbox{K}\to \pi/2$, and the frequency $\omega(E)$ reduces to the
previously obtained quantity $\omega_0$ of Eq.\  (\ref{om0}).
Near the top of the barrier, $P\to 0$, one has $r\to 1$ and 
$\omega(E)$ goes logarthmically to zero.
In the case $H_x=0$ the precession frequency $\omega(E)$ can be 
calculated for the original spin model having the energy levels 
$E_m=-Dm^2$ as the energy difference between the neighboring levels.
The latter is given by $\omega_{m,m+1}\cong 2Dm$, which results in
$\omega(E)=2SD\sqrt{P}$, if we identify $P=m^2/S^2$.
The formula (\ref{omeres}) yields for $\tilde h_x=0$ the same result with $S \Rightarrow\tilde S$, which is an immaterial difference for
$S\gg 1$.    
The dependences $\omega(E)$ for different values of $\tilde h_x$ are
shown in Fig.\  \ref{tatb_ome}.
One can see that they have different types for $h_x\geq 1/4$ and
$h_x<1/4$. 
\par

The energy levels $E_n$ in the wells satisfy the 
Bohr-Sommerfeld quantization rule 
(see, e.g., Ref.\ \onlinecite{lanlif3})
%
\begin{equation}\label{BohrSom}
S^{\rm real}(E_n) \equiv 
2\sqrt{2m}\int_{x_1}^{x_2}\!\! dx\; \sqrt{E_n-U(x)}
= 2\pi\left(n+\frac 12\right)
\end{equation}
and can be found numerically.
The total number of levels in one well is generally given by 
%
\begin{equation}\label{Nlevdef}
N_{\rm lev} = S^{\rm real}(U_{\rm sad})/(2\pi) 
\end{equation}
and in the unbiased case has the explicit form
%
\begin{equation}\label{nlev}
N_{\rm lev} \cong \frac{ 2 S }{ \pi } 
\left[ 
\arctan\sqrt{\frac{ 1- h_x }{ h_x } } - 
\sqrt{ h_x(1- h_x)} 
\right] .
\end{equation}
In the limit $ h_x\to 0$ one has $N_{\rm lev} \cong S$. 
This means that
one half of the total $2S+1\cong 2S$ spin levels are in one well and
the other half are in the other well, thus all the levels above the 
top of the barrier are unphysical (cf. the end of Sec. \ref{mapping}).
In the case $1-h_x \ll 1$ Eq.\  (\ref{nlev}) simplifies to
%
\begin{equation}\label{nlev1}
N_{\rm lev} \cong  (4S/3\pi)(1-h_x)^{3/2} . 
\end{equation}
It can be seen that the quantity $v$ of Eq.\  (\ref{vdef}) 
underestimates the number of levels in the well by factors of 2 at
$h_x\to 0$ and $16/(3\pi)\approx 1.70$ at $h_x\to 1$.

With the imaginary-time action $S(E)$ and the oscillation frequency
$\omega(E)$ having been determined, the problem of finding the level
splitting $\Delta E_n$ of Eq.\  (\ref{splt}) is solved.
In particular, for energies near the bottom of the potential one has $\omega(E)\cong \omega_0$, and $S(E)$
is given by Eq.\  (\ref{sebot}) with $E_n-U_{\rm min}=(n+1/2)\omega_0$.
The level splitting (\ref{splt}) multiplied by the factor 
(\ref{weicor}) takes on the form \cite{she83,weihae83,zas90pla}
%
\begin{equation}\label{spltbot}
\Delta E_n \cong \Delta E_0 \, q^n/n! ,
\end{equation}
where $q$ is given by Eq.\  (\ref{qdef}) and $\Delta E_0$ is the 
ground-state splitting 
%
\begin{equation}\label{splt0}
\Delta E_0 = \frac{ 8\tilde S^{3/2}D }{ \pi^{1/2} }
\left[
\frac{ \exp\sqrt{1-\tilde h_x^2} }{ 1+\sqrt{1-\tilde h_x^2} }
\right]^{2\tilde S} \!\!\!\! (1-\tilde h_x^2)^{5/4} \tilde h_x^{2S}.
\end{equation}
The factor $\tilde h_x^{2S}$ above signals that the ground-state
splitting arises, minimally, in the $2S$th order of a perturbation
theory in the transverse field. 
\cite{korshe78,gar91jpa,harbou95}
More generally, for the excited states one obtains from 
Eq.\ (\ref{spltbot}) $\Delta E_n \propto h_x^{2(S-n)}$. 
\cite{gar91jpa,harbou95}
In our case $S\gg 1$ one can go over from $\tilde S=S+1/2$ and 
$\tilde h_x \equiv h_xS/\tilde S$ to $S$ and $h_x$.
This is not an innocent procedure and it could, in principle, 
change the prefactor in Eq.\  (\ref{splt0}).
However, the corresponding correction terms cancel each other, 
and one arrives at the same expression without tilde.
The latter is a particular case of the result obtained  by Enz and Schilling \cite{enzsch86} for a more general biaxial model with the transverse field (see also Ref.\ \onlinecite{zas90pla}).
Formula (\ref{splt0}) can be rewritten in a more compact form
%
\begin{equation}\label{splt01}
\Delta E_0 =
\frac{ \omega_0 }{ \pi } \sqrt{2\pi q} 
\exp\left[-\frac{S(U_{\rm min}) }{ 2 }\right] ,
\end{equation}
which looks very similar to the starting formula (\ref{splt}).
The extra factor $ \sqrt{2\pi q} \propto S^{1/2}/h_x$ here appears
because in Eq.\  (\ref{splt01}) the action (\ref{sumin}) corresponds
to the bottom of the potential, whereas in Eq.\  (\ref{splt})
the action corresponds to the ground-state energy level, $E_0$. 
This difference due to the zero-point motion between the two actions,
which is described by Eq.\  (\ref{sebot}) is small, but it strongly
affects the prefactor in Eq.\  (\ref{splt01}).

\subsection{Escape rate in the exponential approximation}
\label{exponential}

For $T\ll \Delta U$ the dominant contribution to the integral
(\ref{tst1}) is due to the narrow region of energy $E$ where the 
product $W(E)\exp(-E/T)$ attains its maximum.
Neglecting the prefactor, one can write
%
\begin{equation}\label{gammin}
\Gamma \sim \exp(-F_{\rm min}/T),
\end{equation}
where $F_{\rm min}$ is the minimal value of the effective 
``free energy'' \cite{chugar97}
%
\begin{equation}\label{feff}
F = E + TS(E) - U_{\rm min}
\end{equation}
with respect to $E$.
Since with the exponential accuracy one can set $S(E)=0$ for
$E>U_{\rm sad}$, the free energy $F(E)$ has a downward cusp at
$E=U_{\rm sad}$. 
Thus the minimum of $F(E)$ is attained on the interval
$U_{\rm min} \leq E \leq U_{\rm sad}$. 
The classical regime corresponds, clearly, to the minumum just at the
top of the barrier, $E=U_{\rm sad}$. 
In the quantum regime the energy of the minimum shifts down
and can be found from the condition 
%
\begin{equation}\label{extrcond}
\tau(E) = 1/T,
\end{equation}
where $\tau(E)=-dS(E)/dE$ is the period of oscillations in the 
inverted potential $-U(x)$.
This condition is familiar from quantum statistics. \cite{fey72,aff81,larovc83}
In the instanton language, it determines the trajectory that 
dominates the transition rate at temperature $T$.

\begin{figure}[t]
\unitlength1cm
\begin{picture}(11,7)
\centerline{\epsfig{file=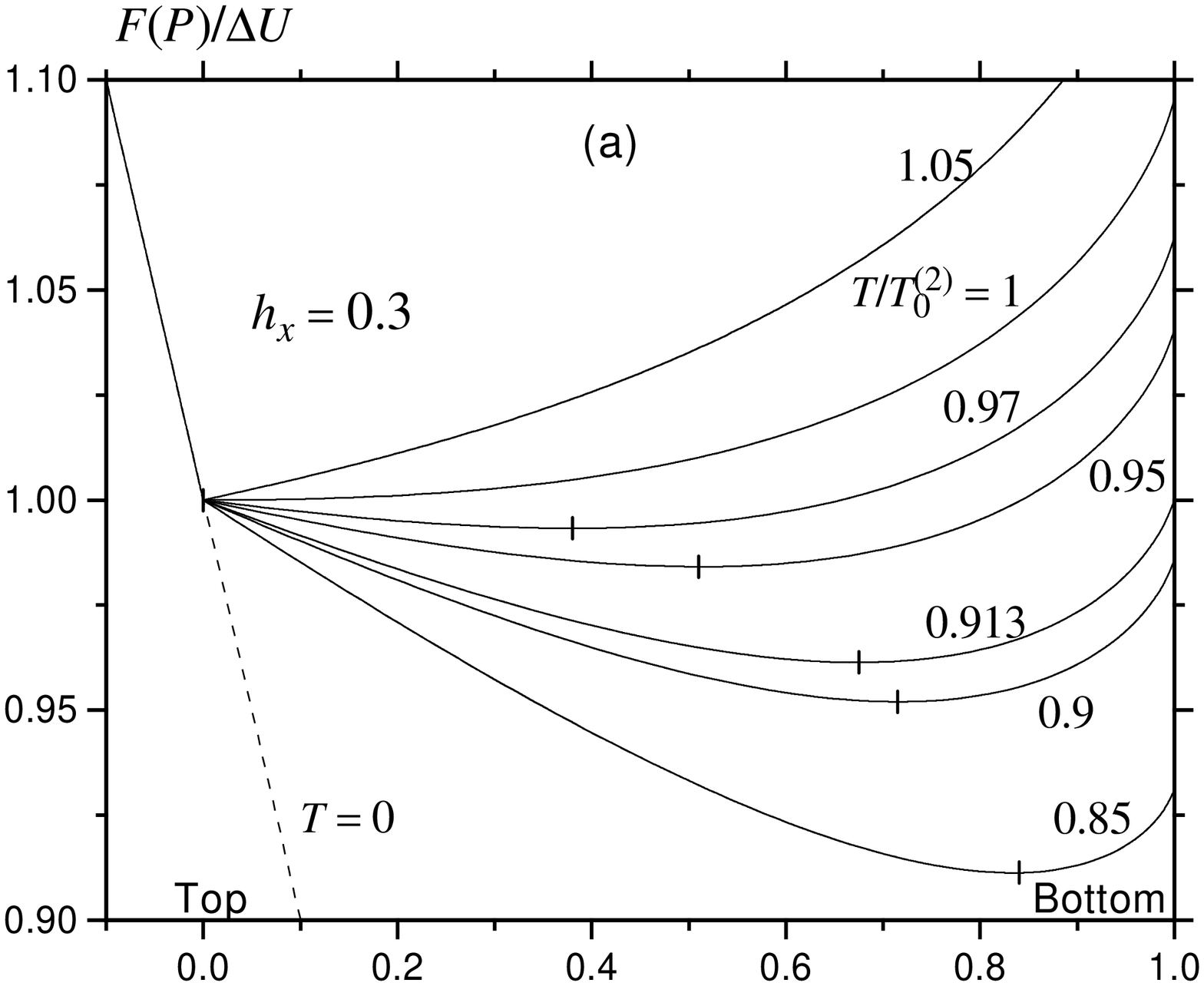,angle=-90,width=12cm}}
\end{picture}
\begin{picture}(11,6)
\centerline{\epsfig{file=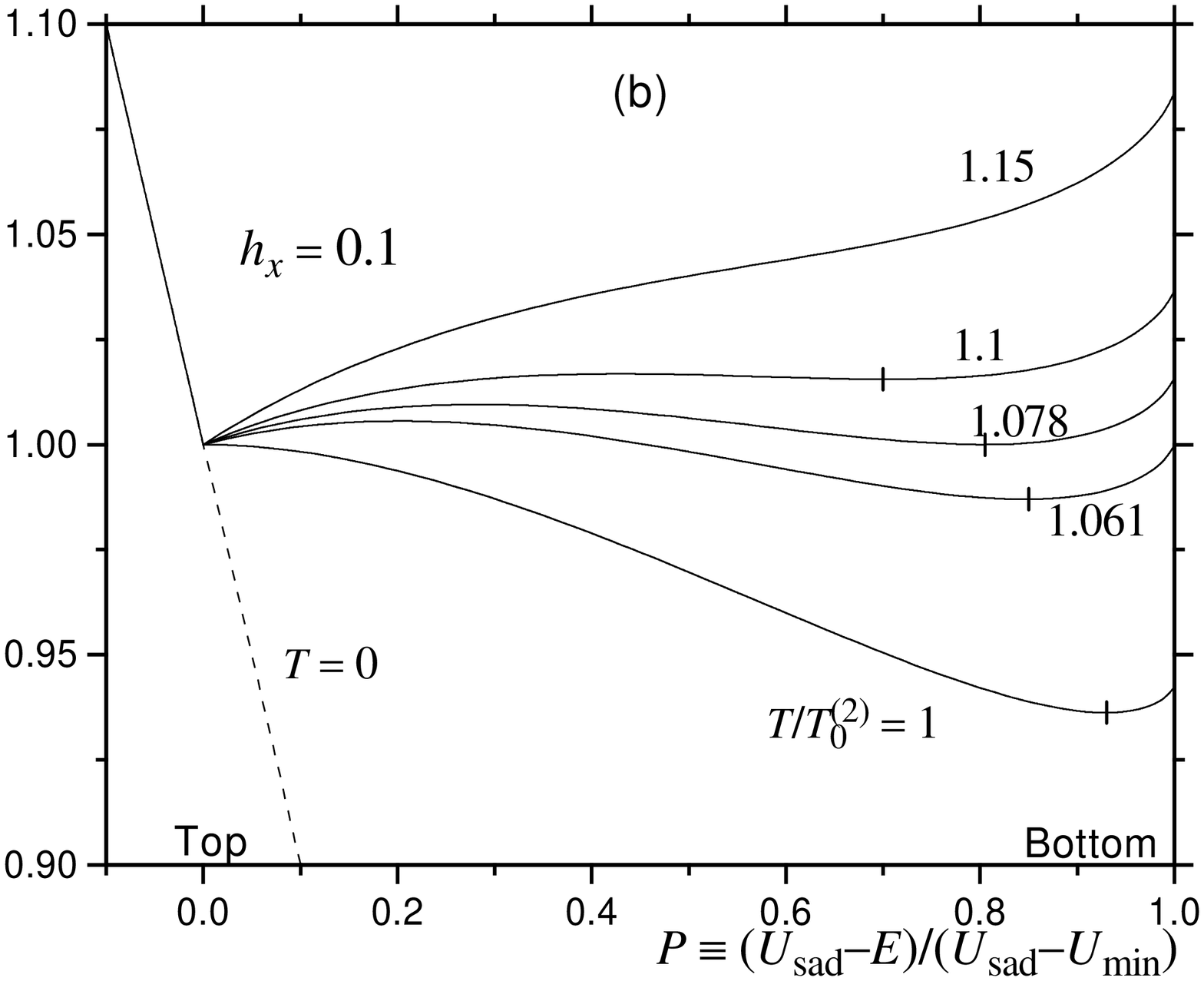,angle=-90,width=12cm}}
\end{picture}
\caption{ \label{tatb_fe} 
Effective ``free energy'' for the escape rate:
(a) -- $h_x = 0.3$, second-order transition;
(b) -- $h_x = 0.1$, first-order transition.
}
\end{figure}

Above the top of the barrier, $P<0$, the effective free energy
(\ref{feff}) has the trivial form [see Eq.\  (\ref{pdef})]
%
\begin{equation}\label{fpup}
F(P) = (1-P)\Delta U = - 2\pi\tilde v T_0^{(2)}P + \Delta U ,
\end{equation}
whereas just below the top of the barrier, $0\leq P\ll 1$, 
it can with the help of Eq.\  (\ref{setop}) be written as
%
\begin{equation}\label{fpsmall}
F(P) \cong 2\pi\tilde v T [ a P + b P^2 + c P^3 + O(P^4)] + \Delta U . 
\end{equation}
Here $a=(T-T_0^{(2)})/T $ with
%
\begin{equation}\label{t02res}
T_0^{(2)} = (SD/\pi)\sqrt{h_x(1-h_x)}  
\end{equation}
[see Eqs.\  (\ref{t02}) and (\ref{tilom0})],
the coefficients $b$ and $c$ are given by Eq.\  (\ref{bcdef}),
and $\tilde v$ is given by Eq.\  (\ref{vtildef}).
The analogy with the Landau model of phase transitions
\cite{lan37}
described by $F = a\phi^2 + b\phi^4 + c\phi^6 + F_0$, now becomes apparent.
The factor $a$ changes
sign at the phase transition temperature $T=T_0^{(2)}$.
The factor $b$ changes sign
at the field value $h_x=1/4$ determining the boundary
between the first- and second-order transitions.
The factor $c$ remains always positive.
The dependence of $F$ on $P$ for the entire range of energy
is plotted in Fig.\  \ref{tatb_fe} with the use of the general analytical expression for $S(E)$, Eq.\  (\ref{seres}).   
At $h_x=0.3$ (Fig.\  \ref{tatb_fe}a) the minimum of $F$ remains 
$\Delta U$ for all $T>T_0^{(2)}$. 
Below $T_0^{(2)}$ it continuously shifts from
the top to the bottom of the potential as temperature is lowered.
This corresponds to the second-order transition from thermal
activation to thermally assisted tunneling, the quantity $P$ playing
the role of the order parameter.
At $h_x=0.1$ (Fig.\  \ref{tatb_fe}b), however, there can be one or two minima of $F$, depending on temperature.
The transition between classical and quantum regimes occurs when
the two minima have the same free energy, which for $h_x=0.1$ takes place at $T_0=1.078 T_0^{(2)}$.

One can see that the criterion of the second-order escape-rate transition is the positiveness of the second derivative of the action
$S(E)$ and, hence, the effective free energy $F(E)$ defined by 
Eq.\  (\ref{feff}).
Because of the relation
%
\begin{equation}\label{tauder}
\frac{ d\tau }{dE} = -\frac{d^2 S}{dE^2} = 
- \frac 1T \frac{d^2 F}{dE^2}
\end{equation}
this criterion is equivalent the requirement that the instanton period $\tau$ monotonically increases with decreasing energy. \cite{chu92}

\begin{figure}[t]
\unitlength1cm
\begin{picture}(11,7)
\centerline{\epsfig{file=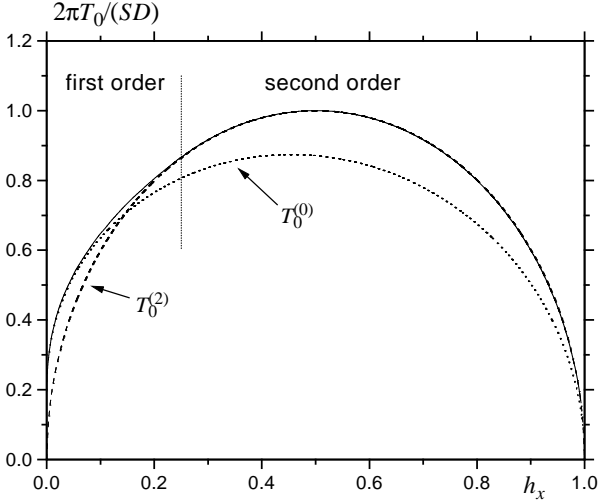,angle=-90,width=12cm}}
\end{picture}
\caption{ \label{tatb_t0} 
Dependence of the crossover temperature $T_0$ on the
transverse field.
}
\end{figure}

\begin{figure}[t]
\unitlength1cm
\begin{picture}(11,7)
\centerline{\epsfig{file=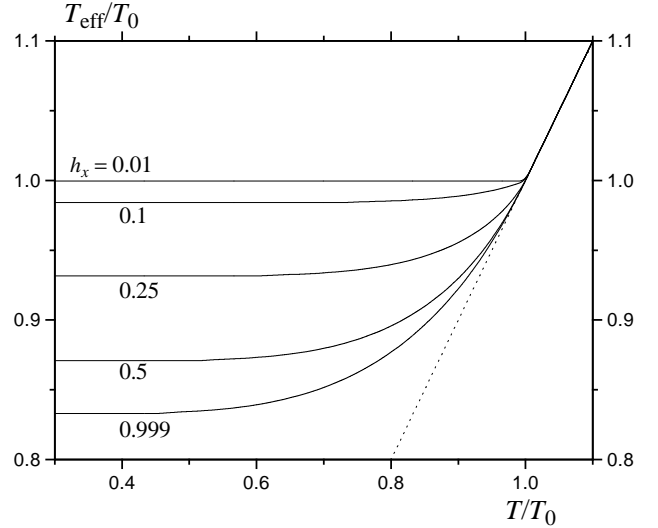,angle=-90,width=12cm}}
\end{picture}
\caption{ \label{tat_f3} 
Dependences of the effective temperature $T_{\rm eff}$ on $T$
for different values of the transverse field.
}
\end{figure}

The ``simple'' estimation for the crossover temperature $T_0$ 
given by Eq.\  (\ref{t0up0}) can with the use of Eq.\  (\ref{deltau})
and $B=S(U_{\rm min})$ of Eq.\  (\ref{sumin}) be explicitly written as 
%
\begin{eqnarray}\label{t00}
&&
T_0^{(0)} = \frac{SD}{4} \frac{ (1-h_x)^2 }
{ \ln\left( \displaystyle \frac{1+\sqrt{1-h_x^2}}{h_x}\right) - 
\sqrt{1-h_x^2} } 
\nonumber\\
&&
\renewcommand{\arraystretch}{2}
\qquad
{}\cong \frac{SD}{4}
\left\{
\begin{array}{ll}
\displaystyle
\frac{1}{\ln[2/(eh_x)]},         
                 & h_x\ll 1  \\
\displaystyle
\frac{3}{2^{3/2}}(1-h_x)^{1/2} ,
                 & 1-h_x\ll 1 .    
\end{array}
\right. 
\end{eqnarray}
One can see from Fig.\  \ref{tatb_fe}b that $T_0^{(0)}$ 
underestimates the crossover temperature.
For $h_x=0.1$ one has $T_0^{(0)}=1.061 T_0^{(2)}<T_0$.
The estimation $T_0^{(0)}$ becomes, however, accurate in the
limit of small $h_x$.
The dependence of the crossover temperature $T_0$ on the
transverse field in the whole range, $0<h_x<1$, is presented in
Fig.\  \ref{tatb_t0}. 
The temperature dependence of the escape rate can be
conveniently written in the form 
$\Gamma \sim \exp(-\Delta U/T_{\rm eff} )$,
where the dependence of 
$T_{\rm eff}\equiv T\Delta U/F_{\rm min}$ on $T$ is presented 
in Fig.\  \ref{tat_f3}  for different $h_x$.
It can be seen from Fig.\  \ref{tat_f3} that the most significant difference between the estimation $T_0^{(0)}$ and the
actual transition temperature $T_0$ arises in the limit of a small
barrier, that is, for $h_x\to 1$.
The former is described by the intersection of the dotted Arrhenius
line with the horizontal line corresponding to the value of 
$T_{\rm eff}(T)/T_0$ at $T=0$. 
From Eqs.\  (\ref{t02}) and (\ref{t00}) for $h_x\to 1$ one obtains 
$T_0^{(0)}/T_0^{(2)}=3\pi/(8\sqrt{2})\approx 0.833$.

The first-order escape-rate transition considered above is the 
transition from thermal activation to thermally assisted tunneling near
the bottom of the potential and {\em not directly} to the ground-state
tunneling. 
This is due to the logarithmic divergence of the instanton period
$\tau$ for the energies near $U_{\rm min} $.
In some field-theoretical models, as, e.g., the reduced nonlinear
$O(3)$-$\sigma$ model, $\tau$ approaches 0 near the bottom of the potential.
Accordingly, the second derivative of $S(E)$ and $F(E)$ is negative
everywhere, as for the rectangular potential for particles.
In such a situation, as it is clear from Fig.\  \ref{tatb_fe}b, the 
minimun of $F(E)$ can only be at $E=U_{\rm sad}$ or $E=U_{\rm min}$.
That is, thermal activation competes directly with the ground-state
tunneling, and the estimation $T_0^{(0)}$ for $T_0$ is exact.
Field theories showing this extreme case of the first-order 
escape-rates transition were called ``type-II theories'', in contrast
to the ``type-I theories'' showing a second-order transition.
In Ref.\ \onlinecite{zimtchmue97} it was shown that adding 
a small Skyrme term to the reduced nonlinear $O(3)$-$\sigma$ 
model causes $\tau$ to diverge near the bottom of the potential,
with the accordingly small amplitude.
This is, in a sense, similar to the situation realized in our spin
model for very small $h_x$ (see Fig.\  \ref{tatb_tau}).  

\par

\subsection{Beyond the exponential approximation}
\label{beyond}

To obtain the escape rate $\Gamma$ with prefactor, one should perform
integration in Eq.\  (\ref{tst1}) or, at lower temperatures, summation
in Eq.\  (\ref{tst}).
This is of a general interest in the situation where the second-order
quantum-classical escape-rate transition is realized in the
exponential approximation. 
Here the effect of thermal distribution leads to quantum corrections
in the classical region of temperature and it smoothens the 
transition to a usual crossover without any singularities of 
$\Gamma(T)$.

In the case of a second-order transition, i.e., $b>0$ in 
Eq.\  (\ref{fpsmall}), there are for $S\gg 1$ four overlapping
temperature ranges
where analytical expressions for $\Gamma(T)$ are available.
Above $T_0^{(2)}$, which exactly means 
$a\gg [b/(2\pi\tilde v)]^{1/2} \sim S^{-1/2}$ 
[see Eqs.\  (\ref{fpsmall}) and (\ref{vtildef})],
one can neglect the terms with $b$ and $c$ in Eq.\  (\ref{setop}) and
extend the integration in Eq.\  (\ref{tst1}) to $\pm\infty$.
This leads to the expression \cite{aff81}
%
\begin{equation}\label{gam1}
\Gamma \cong \frac{ \tilde\omega_0 }{2\pi } 
\frac{ \sinh[\omega_0/(2T)] }{ \sin[\tilde\omega_0/(2T)] }
\exp\left(-\frac{\Delta U}{T}\right) ,
\end{equation}
having the asymtote
%
\begin{equation}\label{gam1as}
\Gamma \cong \frac{ \omega_0 }{2\pi } 
\left( 1 + \frac{ \omega_0^2 + \tilde\omega_0^2 }{ 24 T^2 }\right)  
\exp\left(-\frac{\Delta U}{T}\right) ,
\end{equation}
for $T\gg T_0^{(2)}$, i.e., for $a\gg 1$.
For $T$ approaching $T_0^{(2)}$ from above the prefactor in 
Eq.\  (\ref{gam1}) diverges because of the unlimited contribution from
the range $E<U_{\rm sad}$, i.e., $P>0$, into the integral in 
Eq.\  (\ref{tst1}).

In the temperature range $-b \ll a \ll 1$, i.e., across the transition
region, one can neglect the contribution of the states with 
$E>U_{\rm sad}$ in Eq.\  (\ref{tst1}) and write it in the form
%
\begin{equation}\label{gam2}
\Gamma \cong \frac{ \Delta U e^{-2\pi\tilde v} }{2\pi Z_0 } 
\int_0^\infty \!dP\, \exp[-2\pi\tilde v (aP+bP^2+cP^3)] ,
\end{equation}
where $2\pi\tilde v\equiv\Delta U/T_0^{(2)}$.
If $b$ is not close to zero, one can set $c=0$ and obtain the result
\cite{aff81}
%
\begin{equation}\label{gamerf}
\Gamma \cong \frac{ \tilde\omega_0 }{ 2\pi Z_0 }
\sqrt{\frac{ \tilde v }{ 2b } } 
\exp\left[ 
-\frac{ \Delta U }{ T_0^{(2)} } \left( 1 - \frac{ a^2 }{ 4b } \right) 
\right]  
{\rm erfc} \left( a\sqrt{ \frac{ \pi\tilde v }{ 2b } } \right) .
\end{equation}

For lower temperatures $a \lsim -b$, i.e., $T\lsim T_0^{(2)}$,
one can no longer use the
expansion of $S(E)$ or $F(E)$ near the top of the barrier, but in
this case the integral in Eq.\  (\ref{tst1}) is dominated by its 
stationary point.
One can extend, again, the integration range to $\pm\infty$ and
obtain the result \cite{aff81}
%
\begin{equation}\label{gam3}
\Gamma \cong \frac{ 1 }{ Z_0 }
\frac{ 1 }{ \sqrt{ 2\pi |d\tau/dE| } } 
\exp\left( -\frac{ F_{\rm min} }{ T } \right) , 
\end{equation}
where $F_{\rm min}$ and $d\tau/dE$ are determined by Eqs.\  (\ref{feff}) and (\ref{extrcond}) and can be calculated numerically.  

It can be seen that Eq.\  (\ref{gamerf}) describes the crossover 
between Eqs.\  (\ref{gam1}) and (\ref{gam3}) in the narrow region
$\Delta a \sim [2b/(\pi\tilde v)]^{1/2} \sim S^{-1/2}$ 
around $a=0$.
In this region the erfc function in  Eq.\  (\ref{gamerf}) changes from
2 below $T_0^{(2)}$ to small values above $T_0^{(2)}$.
One can check that the main part of Eq.\  (\ref{gamerf}), except for the 
erfc function, is the concrete form of Eq.\  (\ref{gam3}) in the 
temperature region just below $T_0^{(2)}$.
Thus, Eq.\  (\ref{gam3}) can be extended up to $T_0^{(2)}$ by 
multiplying it by the erfc function.  
It cannot however, be extended above $T_0^{(2)}$ since in this 
region $F_{\rm min}$ in Eq.\  (\ref{gam3}) does not have the same form
as its equivalent in Eq.\  (\ref{gamerf}).
At the boundary between first- and second-order transitions one has
$b=0$, and the term with $c$ in Eq.\  (\ref{gam2}) should be taken
into account.
This leads to qualitatively similar results; the crossover between
Eqs.\  (\ref{gam1}) and (\ref{gam3}) occurs in a narrower region 
$\Delta a \sim c^{1/3}/(2\pi\tilde v)^{2/3} \sim S^{-2/3}$. 
In the range of transverse fields corresponding to the first-order 
escape-rate transition the width of the crossover between the 
classical and quantum regimes is even narrower:
$\Delta a \sim B^{-1} \sim S^{-1}$ 
[see the discussion after Eq.\  (\ref{t0up0})].    
 
In the range of the lowest temperatures one should take into account
quantization of levels in the well and use Eq.\  (\ref{tst}) where
summation runs near the bottom of the potential.
Using the oscillator energy levels 
$E_n = (n+1/2)\omega_0 + U_{\rm min}$ and
Eqs.\  (\ref{gamspl}) and (\ref{spltbot}), one can write  
%
\begin{equation}\label{gam4}
\Gamma \cong (1-e^{-\omega_0/T}) 
\frac{ \pi (\Delta E_0)^2 }{ 2\omega_0 }
\sum_{n=0}^\infty \frac{ [qe^{-\omega_0/(2T)}]^{2n} }{ (n!)^2 } ,
\end{equation}
where the sum is the modified Bessel function $I_0$.
With the help of Eq.\  (\ref{splt01}) the result can be put into the
final form
%
\begin{equation}\label{gam4bes}
\Gamma \cong q\omega_0 (1-e^{-\omega_0/T}) e^{-S(U_{\rm min})}
I_0[2qe^{-\omega_0/(2T)}] .
\end{equation}
Using the asymptotic formula $I_0(x)\cong e^x/\sqrt{2\pi x}$ for
$x\gg 1$ one can check that Eq.\  (\ref{gam4bes}) goes over with raising temperature to Eq.\  (\ref{gam3}) with the parameters calculated from 
the action (\ref{sebot}).
The argument of the Bessel function in Eq.\  (\ref{gam4bes}) is of order
unity for $T\sim T_{00} $, where
%
\begin{equation}\label{t00gro}
T_{00} = \frac{ \omega_0 }{ 2\ln q } 
= \frac{ SD (1-h_x^2)^{1/2} }{ \ln[8S(1-h_x^2)^{3/2}/h_x^2] } 
\end{equation}
[cf. Eq.\  (6.1) of Ref.\ \onlinecite{garchu97}].
The temperature $T_{00}$ characterizes the crossover from thermally
assisted tunneling to the ground-state tunneling; 
for $T\lsim T_{00}$ Eq.\  (\ref{gam4bes}) yields 
$\Gamma_0$ of Eq.\  (\ref{tst}), multiplied by the correction factor
(\ref{weicor}) squared.
For $S\gg 1$ the crossover temperature $T_{00}$ is lower than $T_0$
given by Eq.\  (\ref{t02res}) or Eq.\  (\ref{t00}) because of $S$ under the logarithm.

\begin{figure}[t]
\unitlength1cm
\begin{picture}(11,7)
\centerline{\epsfig{file=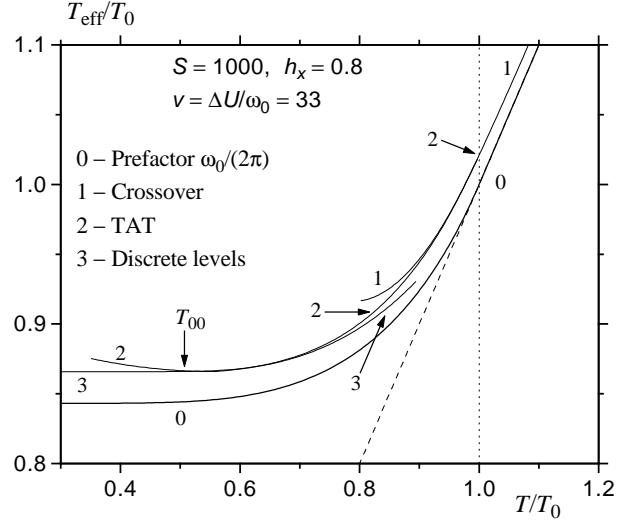,angle=-90,width=12cm}}
\end{picture}
\caption{ \label{tatb_tef} 
Effective temperature for the escape rate in the exponential 
approximation and beyond.
Curve 1 uses Eqs.\  (\protect\ref{gam1}) and (\protect\ref{gamerf})
and describes the crossover from the classical regime with quantum
corrections to thermally assisted tunneling (TAT).
Curve 2 uses  Eq.\  (\protect\ref{gam3}) and describes the TAT regime
in the continuous-level approximation.
Curve 3 uses Eq.\  (\protect\ref{gam4}) for the discrete levels and
describes the crossover from TAT to the ground-state tunneling.
}
\end{figure}
\par
\begin{figure}[t]
\unitlength1cm
\begin{picture}(11,7)
\centerline{\epsfig{file=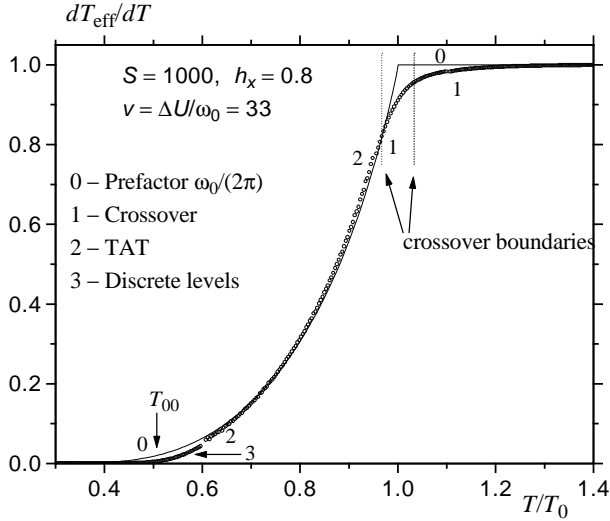,angle=-90,width=12cm}}
\end{picture}
\caption{ \label{tatb_td} 
Derivative of the effective temperature for the escape rate in 
the exponential approximation and beyond.
The small circles represent the numerical derivative of the 
$T_{\rm eff}(T)$ curve obtained by joining the curves 1, 2, and 3
in the figure above.
The crossover boundaries correspond to the argument of the erfc function in Eq.\  (\protect\ref{gamerf}) taking the values $\pm 1$.
}
\end{figure}
\par

The results obtained above can be conveniently represented in terms
of the effective temperature $T_{\rm eff} $ defined by
%
\begin{equation}\label{teffdef}
\Gamma = \frac{ \omega_0 }{2\pi } Q 
\exp\left(-\frac{ F_{\rm min} }{ T }\right)
= \frac{ \omega_0 }{2\pi }
\exp\left(-\frac{ \Delta U }{ T_{\rm eff} }\right),
\end{equation}
where the factor $Q$ accounts for the deviation of the prefactor
from that of the simple TST.
One has, explicitly,
%
\begin{equation}\label{teffres}
T_{\rm eff} = T \Delta U/(F_{\rm min} - \ln Q) .
\end{equation}
The dependence $T_{\rm eff}(T)$ is represented in the situation of
the second-order transition in Fig.\  \ref{tatb_tef}.
One can see that all the analytical curves smoothly join each other.
The escape rate with the accurate prefactor is always higher than
that in the exponential approximation.
For temperatures above the quantum-classical transition this is due
to the nonvanishing quantum transparency of the barrier.
At zero temperature the escape rate is higher because tunneling 
occurs from the ground-state level which is slightly above the bottom
of the potential. 
The derivative $dT_{\rm eff}/T$ is represented in Fig.\  \ref{tatb_td}
where the smoothening of the second-order escape-rate transition
beyond the exponential approximation can be clearly seen.

\section{The biased case}
\label{biased}

In the general biased case, $H_z\ne 0$, the imaginary- and 
real-time actions of Eqs.\  (\ref{action}) and (\ref{BohrSom})
can be still expressed in terms of elliptic integrals, since
$U(x)$ of Eq.\  (\ref{ux}) is proportional to the fourth-order
polynom in $y=\exp(x)$.
We will not do it here because the turning points $x_{1,2}(E)$ for the
motion in the potentials $\pm U(x)$, as well as the extrema of 
$U(x)$, are given by the solution of the fourth-degree algebraic
equation and have a cumbersome analytical form.   
In this section we also restrict ourselves to the exponential 
approximation, since the effects associated with the prefactor
of the escape rate $\Gamma$ do not differ qualitatively from those
analyzed in the unbiased case above.
We will therefore neglect the difference between $\tilde S$ and $S$,
as well as between $\tilde h_{x,z}$ and $h_{x,z}$ here.

\begin{figure}[t]
\unitlength1cm
\begin{picture}(11,7)
\centerline{\epsfig{file=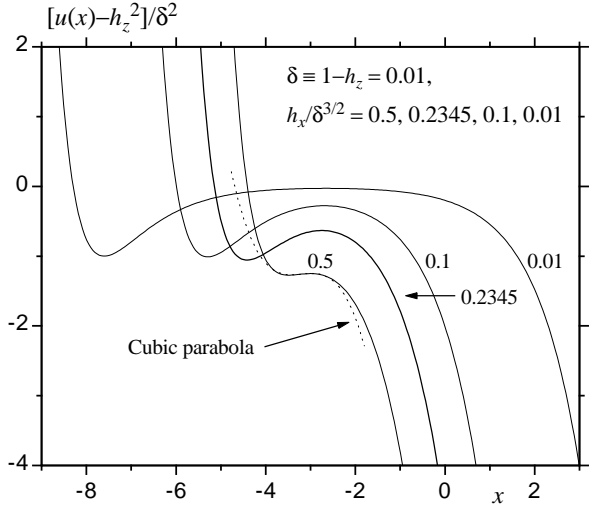,angle=-90,width=12cm}}
\end{picture}
\caption{ \label{tatb_uz} 
Reduced potential $u(x)$ of Eq.\  (\protect\ref{uxz}) for different
values of $h_x$ and $\delta\equiv 1-h_z \ll 1$.  
The value $h_x=0.2345\delta^{3/2}$ corresponds to the boundary between the first- and second-order escape-rate transitions.
}
\end{figure}

It is convenient to start the qualitative analysis from the 
strongly biased case, $\delta\equiv 1-h_z \ll 1$.
Here the metastability boundary curve of Eq.\  (\ref{metbou}) is given
by
%
\begin{equation}\label{metboubi}
h_{xm} = (1-h_z^{2/3})^{3/2} \cong (2\delta/3)^{3/2}
\cong 0.5443\delta^{3/2}.
\end{equation}
The reduced potential $u(x)$ of Eq.\  (\ref{ux}) simplifies in the region of its metastable minimum and maximum to
%
\begin{equation}\label{uxz}
u(x) \cong h_z^2 + h_x[ (h_x/4)e^{-2x} - \delta e^{-x} -2e^x].
\end{equation}
In the above expression the term with $e^{2x}$ which is responsible
for the formation of the stable minimum of $u(x)$ and is small in the
region of interest was dropped.
It can be checked that the metastable minimum of Eq.\  (\ref{uxz})
disappears for $h_x > h_{xm}$.
For $\epsilon_x\equiv (h_{xm}-h_x)/h_{xm}\ll 1$ the potential 
$u(x)$ can be approximated
by a cubic parabola, which will be done in a more general form below.
This is a standard case, \cite{migchu96} in which the second-order
escape-rate transition takes place.
If, on the contrary, the transverse field is removed, the barrier 
height retains a finite value but tunneling should disappear, which
means that the barrier becomes infinitely thick.
Indeed, for $h_x\ll h_{xm} $ the third-degree algebraic equation
determining the extrema of the potential (\ref{uxz}) simplifies 
to yield
%
\begin{eqnarray}\label{uxzxmm}
&&
y_{\rm min} \cong  h_x/(2\delta)
[ 1 + h_x^2/(2\delta^3) ],
\nonumber\\
&&
y_{\rm sad} \cong (\delta/2)^{1/2}
[ 1 + h_x/(2\delta)^{3/2} ],
\end{eqnarray}
with $y_{\rm min, sad}\equiv \exp(x_{\rm min, sad})$ and
%
\begin{eqnarray}\label{uxzumm}
&&
u_{\rm min}-h_z^2 \cong -\delta^2
[ 1 + h_x^2/\delta^3 ],
\nonumber\\
&&
u_{\rm sad}-h_z^2 \cong -2^{3/2}h_x \delta^{1/2}
[ 1 - h_x/( 2^{5/2}\delta^{3/2}) ].
\end{eqnarray}
One can see that in the limit $h_x\to 0$ the point $x_{\rm sad}$ is
fixed, $x_{\rm min}$ goes to $-\infty$, and the barrier height
%
\begin{equation}\label{delubi}
\Delta U \cong S^2 D \delta^2 
\left[ 1 - \left(\frac 43 \right)^{3/2} \frac{ h_x }{ h_{xm} } 
+ \frac 49 \left( \frac{ h_x }{ h_{xm} } \right)^2 \right] 
\end{equation}
remains finite [cf. Eq.\  (\ref{deltau})].
As in the unbiased case, the very flat top of $U(x)$ 
favors the first-order escape-rate transition.
The solution given by Eqs.\  (\ref{uxzxmm}) and (\ref{uxzumm}) becomes
invalid for $h_x\sim h_{xm}$, where $x_{\rm min}\sim x_{\rm sad}$.
In this region the crossover from first- to second-order transition
is expected. 
The form of the potential $u(x)$ for different values of $h_x$ in
the strongly biased case is shown in Fig.\  \ref{tatb_uz}.
{\sl Its remarkable feature is that both first- and second-order 
transitions are realized for whatever small barrier}, 
$\Delta U \lsim S^2 D \delta^2$, 
{\sl with the arbitrarily small} $\delta$.
This is especially interesting for the experiments on small magnetic
particles, $S\sim 10^5-10^6$, where the barrier should be reduced to
achieve measurable escape rates.  
\par

\begin{figure}[t]
\unitlength1cm
\begin{picture}(11,7)
\centerline{\epsfig{file=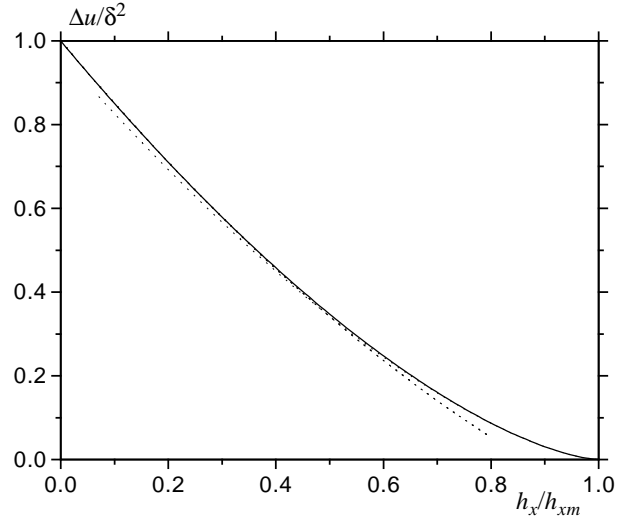,angle=-90,width=12cm}}
\end{picture}
\caption{ \label{tatb_dub} 
The numerically calculated reduced barrier height 
$\Delta u \equiv \Delta U/(S^2D)$ in the 
strongly biased case.
Asymptotes (\protect\ref{delubi}) and (\protect\ref{delucubx}) shown by the dotted lines reproduce the accurate numerical result practically in the whole range of $h_x/h_{xm}$.
}
\end{figure}
\par

\begin{figure}[t]
\unitlength1cm
\begin{picture}(11,7)
\centerline{\epsfig{file=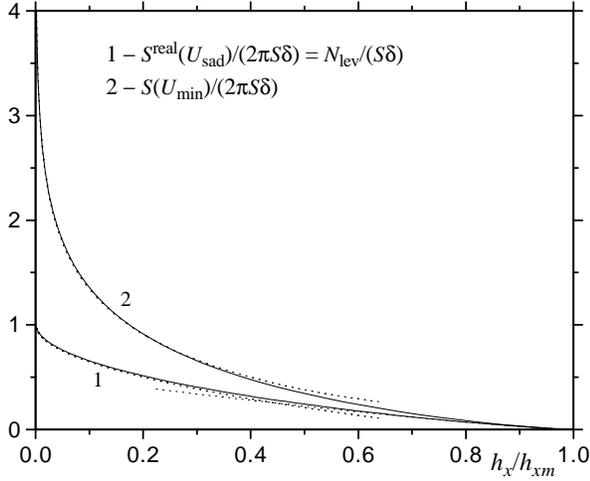,angle=-90,width=12cm}}
\end{picture}
\caption{ \label{tatb_acb} 
Imaginary- and real-time actions in the strongly biased case, Eqs.\  (\protect\ref{suminbiarb}) and (\protect\ref{sumaxbiarb}).
Asymptotes of Eqs.\  (\protect\ref{suminbism}), (\protect\ref{sumaxbism}), and (\protect\ref{suminmax}) are shown by
the dotted lines.  
}
\end{figure}
\par

\begin{figure}[t]
\unitlength1cm
\begin{picture}(11,7.25)
\centerline{\epsfig{file=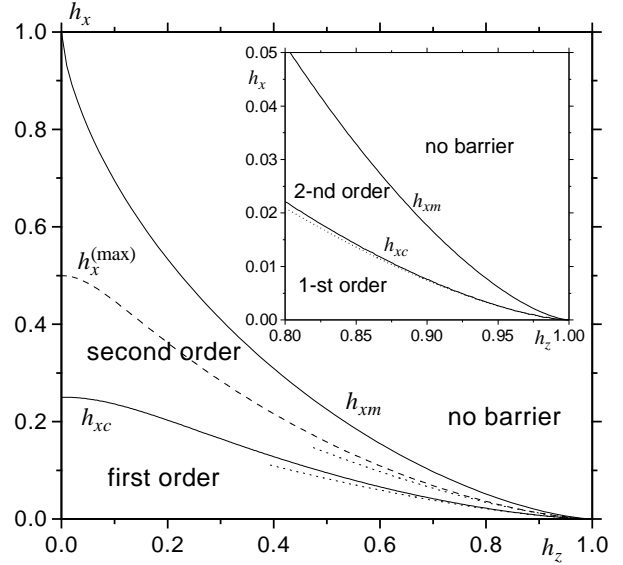,angle=-90,width=12cm}}
\end{picture}
\vspace{0.5cm}
\caption{ \label{tatb_phd} 
Phase diagram for the quantum-classical escape-rate transition.
The dashed line corresponds to the maximum of $T_0$ as function of 
$h_x$.
The asymptotes of Eqs.\  (\protect\ref{hxc}) and
(\protect\ref{t0max}) are shown by the dotted lines.
}
\end{figure}
\par

\begin{figure}[t]
\unitlength1cm
\begin{picture}(11,7)
\centerline{\epsfig{file=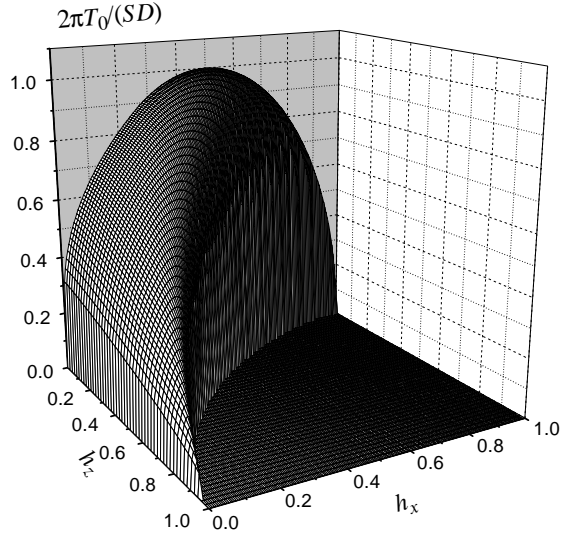,angle=-90,width=12cm}}
\end{picture}
\caption{ \label{tatb_t3d} 
3$d$ plot of $T_0(h_x,h_z)$. In the main part of the interval 
$0 \leq h_z \leq 1$ the transition temperature $T_0$ scales with
$\delta \equiv 1-h_z$.
}
\end{figure}
\par

\begin{figure}[t]
\unitlength1cm
\begin{picture}(11,7)
\centerline{\epsfig{file=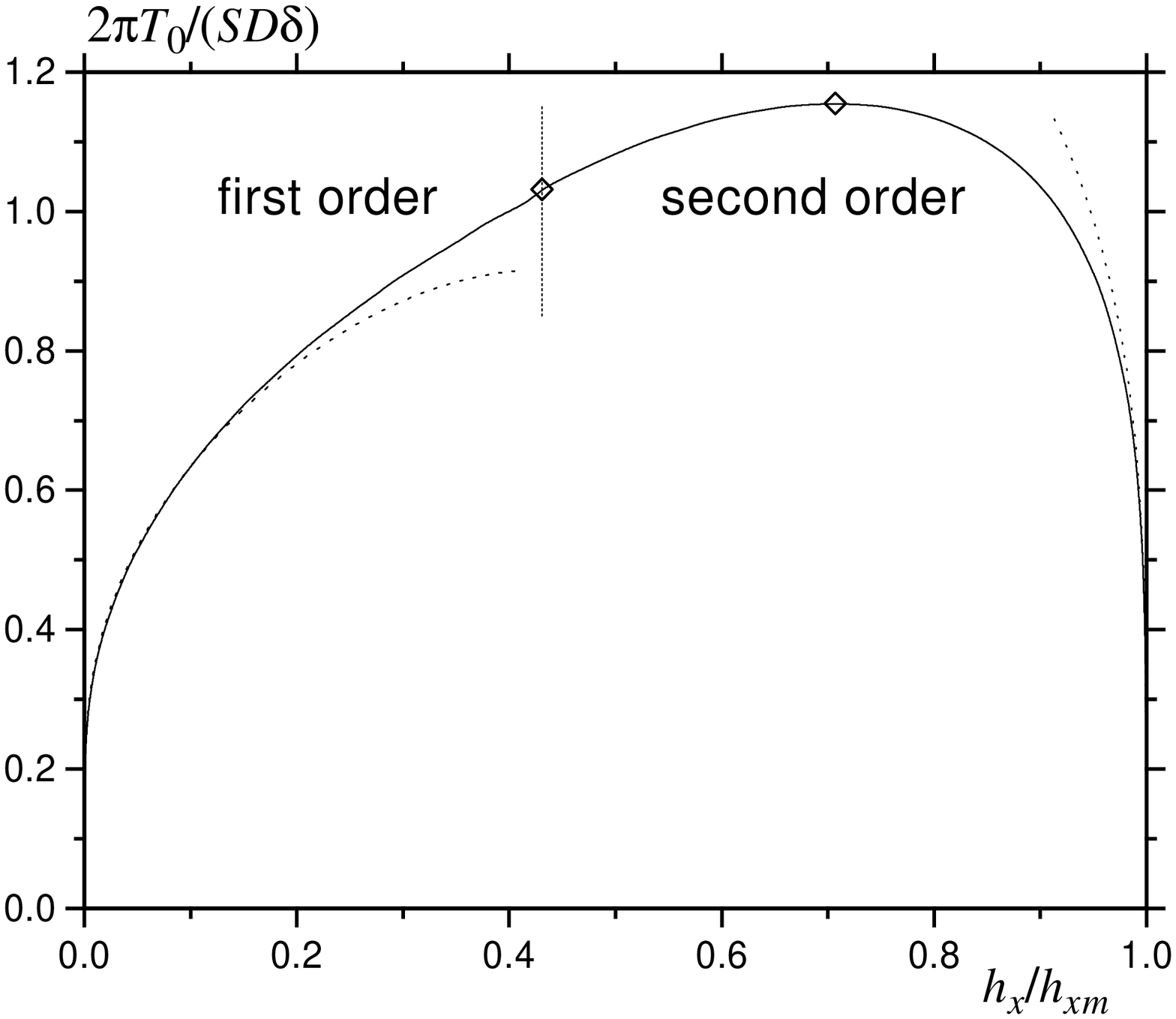,angle=-90,width=12cm}}
\end{picture}
\caption{ \label{tatb_tbi} 
Dependence $T_0(h_x)$ in the strongly biased case 
(cf. Fig.\  \protect\ref{tatb_t0}).
Asymptotes based on Eqs.\  (\protect\ref{t0up0}), 
(\protect\ref{delubi}), and (\protect\ref{suminbism}), on the left
side, and on Eqs.\  (\protect\ref{t02}) and (\protect\ref{ommigchu}),
on the right side, are shown by the dotted lines.
The analytically calculated values of Eqs.\  
(\protect\ref{hxc})--(\protect\ref{t0max}) are shown by diamonds.                
}
\end{figure}
\par

In the general biased case simple analytical results for the 
quantum-classical transition temperature $T_0$ can be obtained in
two limiting cases: for small transverse fields and for the fields
near the metastability boundary curve.
In the first case the quantum-classical transition is of the first
order, and a good estimation of the transition temperature $T_0$ is
given by Eq.\  (\ref{t0up0}), where in the exponential approximation 
$B=S(U_{\rm min})$.
The analytical expression for the bottom-level action 
$S(U_{\rm min})$ can be found for small $h_x$ and arbitrary $h_z$
from the general formula (\ref{action}).
In the most interesting strongly biased case this can be done for the arbitrary $h_x<h_{xm}$ with the result
%
\begin{eqnarray}\label{suminbiarb}
&&
S(U_{\rm min})\cong 2S \sqrt{ 2h_x y_{\rm min}'} 
\big[- 3\sqrt{1-r_{\rm min}}
\nonumber\\
&& 
\qquad\qquad
{} + (r_{\rm min} + 2) \arctanh\sqrt{1-r_{\rm min}}
\big], 
\end{eqnarray}
where $r_{\rm min}\equiv y_{\rm min}/y_{\rm min}'<1$ and 
$y_{\rm min}'\equiv \exp(x_{\rm min}')$ is the turning point on the
right side of the barrier corresponding to the energy $u_{\rm min}$ 
of Eq.\  (\ref{uxzumm}).
For $h_x\ll h_{xm}$ one has 
$y_{\rm min}'\cong [\delta^2/(2h_x)](1-h_x^2/\delta^3 )$ and 
$r_{\rm min} \cong h_x^2/\delta^3\ll 1$.
In this case Eq.\  (\ref{suminbiarb}) simplifies to
%
\begin{equation}\label{suminbism}
S(U_{\rm min})\cong 2S \delta
\left\{ \ln
\left[
\frac{ 27 }{ 2e^3 }\left( \frac{ h_{xm} }{ h_x } \right)^2 
\right] 
+ \frac{ 22 }{ 27 } \left( \frac{ h_x }{ h_{xm} } \right)^2 
\right\} .
\end{equation}

It is instructive to find $S(U_{\rm min})$ for small $h_x$ in the whole range of $h_z$ with the use of the general 
perturbative expression for the level splitttings in the biased case
\cite{chufri,garchu97}
%
\begin{eqnarray}\label{splitbi}
&&
\Delta E_{mm'} = \frac{2D}{[(m'-m-1)!]^2}
\nonumber\\
&&
\qquad
\times \sqrt{\frac{ (S+m')! (S-m)! }{ (S-m')! (S+m)! } }
\left( \frac{ H_x }{ 2D } \right)^{m'-m}
\end{eqnarray}
[cf. Eq.\ (\ref{spltmgen})], where $m<0$ and $m'=-m -H_z/D$ is the
matching level on the other side of the barrier.
For the ground-state level in the metastable well one has $m=-S$ and
$m'=S(1-2h_z)$.
Now, in the exponential approximation, one can use, according to
Eq.\  (\ref{splt}), 
$S^{\rm pert}(U_{\rm min})\cong -2\ln\Delta E_{-S,S(1-2h_z)}$.
Then with the help of the Stirling formula $n!\cong (n/e)^n$ one
arrives at the result
%
\begin{equation}\label{suminbi}
S^{\rm pert}(U_{\rm min})\cong 2S 
\{ \delta \ln[4\delta^3/(e h_x)^2] + h_z\ln h_z \}.
\end{equation}
One can check that in the unbiased case 
($h_z=0$, i.e., $\delta=1$) the
first (perturbative) term of Eq.\ (\ref{suminpert}) is reproduced,
and in the strongly biased case that of Eq.\  (\ref{suminbism}) is
recovered.      
The accuracy of Eq.\  (\ref{suminbi}) is not so high as that
of Eq.\  (\ref{suminbism}), which contains the important correction
term quadratic in $h_x/h_{xm}$.  
Finally, in the strongly biased case the estimation of the 
escape-rate transition temperature for small transverse fields is  
$T_0\cong T_0^{(0)} = \Delta U/S(U_{\rm min})$ with $\Delta U$ and
$S(U_{\rm min})$ given by Eqs.\  (\ref{delubi}) and (\ref{suminbism}).

Near the metastability boundary curve the consideration
begins with the location of the latter from the conditions 
$u'(x_m)=0$ and $u''(x_m)=0$, which yields the equations
%
\begin{equation}\label{shchxm}
\cosh^3(x_m) = \frac{1}{h_{xm}}, 
\qquad \sinh^3(x_m) = -\frac{h_{zm}}{h_{xm}}, 
\end{equation}
where from Eq.\  (\ref{metbou}) follows. 
Then for the field
%
\begin{equation}\label{epsxz}
h_x = h_{xm}(1-\epsilon_x), \qquad h_z = h_{zm}(1-\epsilon_z), 
\end{equation}
where $\epsilon_{x,z}\ll 1$ and $\{h_{xm},h_{zm}\}$ are related by
Eq.\  (\ref{metbou}), one can expand the potential $u(x)$ 
of Eq.\  (\ref{ux}) near $x_m$ in powers of $x-x_m $.
If $h_z$ is not very small, one can restrict oneself to the 
third-order terms and obtain
%
\begin{equation}\label{cubpar}
u(x) \cong u(x_m) +A_1(x-x_m) - A_3(x-x_m)^3
\end{equation}
with
%
\begin{equation}\label{A13}
A_1 = 2\bar\epsilon h_{xm}^{2/3} h_{zm}^{1/3},
\qquad A_3 = h_{xm}^{4/3} h_{zm}^{1/3},
\end{equation}
and 
%
\begin{equation}\label{epsbar}
\bar\epsilon \equiv h_{xm}^{2/3}\epsilon_x + h_{zm}^{2/3}\epsilon_z.
\end{equation}
The cubic parabola (\ref{cubpar}) is symmetric about $x_m$ and it
is characterized by the barrier height
%
\begin{equation}\label{delucub}
\Delta U = 4S^2D (2\bar\epsilon/3)^{3/2} (h_{xm}h_{zm})^{2/3}
\end{equation}
and the equal real and instanton oscillation frequencies
%
\begin{equation}\label{omcub}
\omega_0 = \tilde\omega_0 = 2SD (6\bar\epsilon)^{1/4} 
h_{xm}^{1/2} h_{zm}^{1/6},
\end{equation}
where the latter defines the quantum-classical transition temperature: 
$T_0=T_0^{(2)}=\tilde\omega_0/(2\pi)$.
Equation (\ref{omcub}) can be put into the form 
%
\begin{equation}\label{ommigchu}
\omega_0 = \tilde\omega_0 = 2SD (6\bar\epsilon)^{1/4} 
\cot^{1/6}\theta_H/( 1 + \cot^{2/3}\theta_H),
\end{equation}
if one sets $h_{xm}=h_m\sin\theta_H$ and 
$h_{zm}=h_m\cos\theta_H$, where, by virtue of Eq.\  (\ref{metbou}),
$h_m = (\sin^{2/3}\theta_H + \cos^{2/3}\theta_H )^{-3/2}$.
If, as in Ref.\ \onlinecite{migchu96}, the field changes along the line
$\theta_H=\const$, one has $\epsilon_x=\epsilon_z=\epsilon$, which
results in $\bar\epsilon=\epsilon$.
In this case Eq.\  (18) of Ref.\   \onlinecite{migchu96} is recovered,
where, by definition, $\omega_0$ is twice as small as here. 
In the strongly biased case, however, it is more convenient to have
$h_z$ fixed and swipe $h_x$ across the narrow region where the barrier
exists.
Thus one can set $h_{zm}\cong 1$, $h_{xm} \cong (2\delta/3)^{3/2}$,
$\epsilon_z=0$, and 
$\bar\epsilon = h_{xm}^{2/3}\epsilon_x = (2\delta/3)\epsilon_x$ 
to obtain
%
\begin{equation}\label{delucubx}
\Delta U = (16S^2D/9)(2\epsilon_x/3)^{3/2}\delta^2
\end{equation}
and 
%
\begin{equation}\label{omcubx}
\omega_0 = \tilde\omega_0 = (4SD/3)(6\epsilon_x)^{1/4}\delta. 
\end{equation}
Note that at the applicability boundary of the present aproach,
$\epsilon_x=1$, one has $\Delta U = 0.97 S^2D\delta^2$, which is
very close to the exact value $\Delta U = S^2D\delta^2$ for
$h_x\ll h_{xm}$ obtained above (see Fig.\  \ref{tatb_dub}).
That is, the barrier height scales as $\Delta U \propto \delta^2$
in the strongly biased case $\delta\ll 1$.
Similarly, the quantum-classical transition temperature scales as
$T_0 \propto \delta$. 
In fact, as will be shown by the numerical calculations below, 
this scaling holds not only in the strongly biased case, 
but practically in the whole region of $h_z$, excluding that of very
small $h_z$.

As a pendant to Eq.\  (\ref{suminbiarb}), one can calculate the 
real-time action given by  Eq.\  (\ref{BohrSom}) for the energy 
$U_{\rm sad}$ in the strongly-biased case.
The result reads 
%
\begin{eqnarray}\label{sumaxbiarb}
&&
S^{\rm real}(U_{\rm sad})\cong 2S \sqrt{ 2h_x y_{\rm sad}'} 
\big[- 3\sqrt{r_{\rm sad}-1} 
\nonumber\\
&&
\qquad\qquad
{} + (r_{\rm sad}+2)
\arctan\sqrt{r_{\rm sad}-1}\big] ,
\end{eqnarray}
where $r_{\rm sad}\equiv y_{\rm sad}/y_{\rm sad}'>1$ and 
$y_{\rm sad}'\equiv \exp(x_{\rm sad}')$ is the turning point on the 
left side of the metastable well corresponding to the energy 
$u_{\rm sad}$ of Eq.\  (\ref{uxzumm}).
The top-level real-time action above determines according to 
Eq.\  (\ref{Nlevdef}) the number of levels in the well. 
For $h_x\ll h_{xm}$ one has 
$y_{\rm sad}'\cong h_x/(4\delta)$ and 
$r_{\rm sad} \cong (2\delta)^{3/2}/h_x \gg 1$.
Thus in this limit 
%
\begin{equation}\label{sumaxbism}
S^{\rm real}(U_{\rm sad}) \cong 
2\pi S\delta[1 - 8/(3^{3/4}\pi)\sqrt{ h_x/h_{xm} } ]
\end{equation}
[cf. Eq.\  (\ref{suminbism})]
and $N_{\rm lev}=S\delta$ for $h_x=0$.
Near the metastability boundary the potential $u(x)$ can be approximated by the cubic parabola (\ref{cubpar}).
From Eq.\  (\ref{shchxm}) one obtains 
$y_{\rm min}'\cong y_{\rm sad}'\cong \exp(x_m)= 
(1-h_{zm}^{1/3})/h_{xm}^{1/3}\cong \delta/(3h_{xm}^{1/3})$.
More accurately, 
$x_{\rm sad, min}\cong x_m \pm\sqrt{A_1/(3A_3)} = 
x_m \pm \sqrt{2\epsilon_x/3}$ and 
$x_{\rm sad, min}'\cong x_m \mp 2\sqrt{2\epsilon_x/3}$.
This yields $1-r_{\rm min} =1-\exp(x_{\rm min} -x_{\rm min}')
\cong \sqrt{6\epsilon_x}$, and the same for
$r_{\rm sad}-1$.
Expanding now Eqs.\  (\ref{suminbiarb}) and (\ref{sumaxbiarb}) for 
$\epsilon_x\ll 1$ results in
%
\begin{equation}\label{suminmax}
S(U_{\rm min}) \cong S^{\rm real}(U_{\rm sad}) \cong
(16S\delta/45)(6\epsilon_x)^{5/4}.
\end{equation}
The behavior of $S(U_{\rm min})$ and $S^{\rm real}(U_{\rm sad})$
given by Eqs.\  (\ref{suminbiarb}) and (\ref{sumaxbiarb}) in the 
strongly biased case in the whole range of $h_x/h_{xm}$ is shown in
Fig.\  \ref{tatb_acb}.

The order of the quantum-classical escape-rate transition is
determined, as we have seen, by the sign of the the coefficient $b$
in the expansion of the imaginary-time action $S(E)$ near the top of
the barrier [see Eqs.\  (\ref{setop}) and (\ref{fpsmall})].
In the general biased case this coefficient is given by
%
\begin{equation}\label{bgen}
b = \frac{ \Delta U }{ (2m)^3 2\tilde \omega_0^6 }
\left[ U_{\rm sad}'' U_{\rm sad}'''' - \frac 53 (U_{\rm sad}''')^2
\right]   
\end{equation}
[see Eqs.\  (\ref{t02}) and (\ref{partpro})].
It reduces to that of Eq.\  (\ref{bcdef}) in the case $h_z=0$ where
$U_{\rm sad}'''=0$.
In the strongly biased case the potential of the spin system 
simplifies to Eq.\  (\ref{uxz}), and there is a simple analytical 
solution for the line $h_{xc}(h_z)$ separating
the first- and second-order escape-rate transitions.
To obtain this line, one can calculate all needed derivatives of $u(x)$ at the top of the barrier from Eq.\  (\ref{uxz}), 
equate $b=0$ in Eq.\  (\ref{bgen}),
and eliminate the terms with $h_x$ with the help of the relation
$h_xe^{-2x} = 2(-2e^x+\delta e^{-x})$ following from 
$u_{\rm sad}'=0$.
Then the condition $b=0$ yields $e^{2x}=\delta/\sqrt{6}$, and for 
$h_x=h_{xc}$ one finds
%
\begin{equation}\label{hxc}
h_{xc} \cong(2/3)^{3/4}(\sqrt{3}-\sqrt{2} )\delta^{3/2}\cong 0.2345 \delta^{3/2},  
\end{equation}
i.e., $h_{xc} \cong 0.4308 h_{xm}$. 
Further, for $h_x=h_{xc}$ one has 
$u_{\rm sad}''\cong -2\delta^2(4\sqrt{2/3}-3)$ resulting in the
transition temperature at the boundary between first-and second-order
transitions
%
\begin{equation}\label{t0c}
T_0^{(c)} = (SD/\pi) (4\sqrt{2/3}-3)^{1/2} \delta,
\end{equation}
where $(4\sqrt{2/3}-3)^{1/2}\cong 0.5157$ 
[see Eq.\  (\ref{t02}), cf. Eq.\  (\ref{t02res})].

The maximum of $T_0$ is attained within the range of transverse fields
corresponding to the second-order escape-rate transition.
In the unbiased case Eq.\  (\ref{t02res}) yields evidently
$T_0^{(\rm max)} = SD/(2\pi)$ for $h_x=h_x^{(\rm max)}=0.5$.
In the strongly biased case using the conditions
$d u''[h_x,x(h_x)]/dh_x=0$ and $u'[h_x,x(h_x)]=0$ for $x(h_x)$
corresponding to the maximum of the potential (\ref{uxz}) one obtains
$e^{2x}=\delta/3$ and 
%
\begin{equation}\label{t0max}
T_0^{(\rm max)} = (SD/\pi) \delta/\sqrt{3}
\end{equation}
for $h_x^{(\rm max)}=h_{xm}/\sqrt{2} \cong 0.3849 \delta^{3/2}$.

Numerical calculation of the escape rate $\Gamma$ in the general
biased case in the exponential approximation poses no problems.
The minimum of the effective free energy $F(E)$ of Eq.\  (\ref{feff})
with respect to $E$ can be found using the 
imagimary-time action $S(E)$ numerically calculated from  
Eq.\  (\ref{action}).
For any field ${\bf H}$ one can establish the transition temperature
$T_0$,
such as for $T<T_0$ the minimum of $F(E)$ no longer corresponds to the
top of the barrier, as it is in the classical case.
Analyzing the dependence $F(E)$ for different fields allows one to determine the order of the quantum-classical escape-rate transition (see Fig.\  \ref{tatb_fe}).
The boundary between the first- and the second-order transitions
can be found the most easily from the condition $b=0$ in 
Eq.\  (\ref{bgen}), where the derivatives of $U(x)$ are calculated at
the numerically determined top of the barrier.
This method yields the same results for $h_{xc}(h_z)$ as that described
above.
The resulting phase diagram for the escape-rate transition is 
shown in Fig.\  \ref{tatb_phd}.
The 3$d$ plot of $T_0(h_x,h_z)$ is shown in Fig.\  \ref{tatb_t3d}.
The dependence $T_0(h_x)$ in the strongly-biased case is given in 
Fig.\  \ref{tatb_tbi}.

\section{Discussion}
\label{discussion}

We have presented a comprehensive study of the thermal and quantum decay of a metastable spin state of the uniaxial spin system in the arbitrarily directed magnetic field. 
The moderate damping regime has been studied, in which the damping 
does not influence the dynamics of the spin system but provides the
thermal equilibrium with the environment.
The method employed is the mapping of a spin system onto the particle 
in a double-well potential, with the subsequent use of the WKB 
approach.
The explicit dependence of the escape rate, including the prefactor,
on temperature, field, and anisotropy constant has been worked out and
compared with limiting cases obtained by others.   
This calculation shows how formulas describing different regimes join
on temperature down to lowest temperatures where quantization of 
levels becomes significant.
The crossover from thermally assisted tunneling to the ground-state
tunneling at $T\sim T_{00}$ is described quantitatively and discussed
in detail [see Eqs.\  (\ref{gam4bes}) and (\ref{t00gro})].     
The fascinating new feature of this analysis is the existence for 
spin systems of both, first- and second-order, transitions from
thermal activation to thermally assisted tunneling at $T=T_0$.
The kind of transition depends on the strength and the direction
of the magnetic field.
We have calculated the boundary in the $H_x,H_z$ plane 
separating the two different regimes.

The direct analogy with phase transitions exists in the limit of 
a very large spin $S$.
In that limit the dependence of the transition rate on temperature
changes abruptly at $T_0$, pretty much as thermodynamic quantities do
in the theory of phase transitions.
For finite $S$, both the first- and second-order transitions of the escape rate are smeared, similar to the smearing of the phase transition in a finite-size system.
The reduced width of the crossover between the classical and quantum
regimes $\Delta T/T_0$ is of the order of $S^{-1/2}$ for the 
second-order transition and $S^{-1}$ for the first-order 
transition.

For a moderately large spin, $S\sim 10$, one can explore in experiment 
the entire phase diagram shown in Fig.\  \ref{tatb_phd}.
This is the case of molecular magnets like Mn$_{12}$Ac.
If greater spins are studied, the external magnetic field must be 
adjusted such that the energy barrier becomes small enough to provide
a significant tunneling rate.
In terms of the phase diagram this means that one has to work close to
the metastability boundary line, Eq.\  (\ref{metbou}), separating the
field ranges with and without the barrier.
According to Fig.\  \ref{tatb_phd}, for a uniaxial spin close to the
metastability boundary, both, first- and second-order, transitions
coexist only in the lower right corner of the phase diagram.
In this region, close to the boundary between first- and second-order
transitions, the temperature $T_0$ is of order $BD/(4\pi^2)$, where
$B$ is the exponent in the expression for the rate.
In a typical tunneling experiment with a macroscopic lifetime of a
metastable state, $B\sim 4\pi^2$, so that $T_0\sim D$.
The latter constant can be expressed in terms of the anisotropy
field and the total spin of the particle: $D=g\mu_B H_A/(2S)$.
For $H_A\sim 1$~T and $S\sim 100$ the transition temperature will be
of the order of 10 mK.
This is within experimental reach.
Notice that the smallness of $T_0$ in our model comes, in part, from
the fact that the noncommutation of $S_z$ with the Hamiltonian is
small in the lower right corner of the phase diagram, where the effect
is to be searched for.
One can expect that in models with transverse anisotropy observable
first- and second-order transitions will coexist at higher temperatures, since the transverse anisotropy, rather than the 
required small transverse field, will drive the decay of the 
metastable state.
Such a model requires a different approach and will be worked out
elsewhere.

\section*{Acknowledgment}

This work has been supported by 
the U.S. National Science Foundation through
Grant No. DMR-9024250.



\begin{thebibliography}{10}

\vspace{-1.5cm}


\bibitem[*]{e-gar} Permanent address: I. Institut f\"ur Theoretische Physik, Universit\"at Hamburg,
Jungiusstrasse 9, D-20355 Hamburg, Germany. 

Electronic addresses: garanin@physnet.uni-hamburg.de\\
garanin@mpipks-dresden.mpg.de\\
 http://www.mpipks-dresden.mpg.de/$\sim$garanin/


\bibitem[\dagger]{e-mar}
Permanent address: 
Departament de F\'\i sica Fonamental, Universitat de Barcelona, 
Av. Diagonal 647, 08028 Barcelona, Spain

 Electronic address: xavim@hermes.ffn.ub.es 

\bibitem[\ddagger]{e-chu}
Electronic address: chudnov@lcvax.lehman.cuny.edu\\


\bibitem{hun27}
{F. Hund}, Z. Phys. {\bf 43},  805  (1927).

\bibitem{wen26}
{G. Wentzel}, Z. Phys. {\bf 38},  518  (1926).

\bibitem{kra26}
{H. A. Kramers}, Z. Phys. {\bf 39},  828  (1926).

\bibitem{bri26}
{L. Brillouin}, C. R. Acad. Sci. Paris {\bf 183},  24  (1926).

\bibitem{opp28}
{J. R. Oppenheimer}, Phys. Rev. {\bf 31},  80  (1928).

\bibitem{fownor28}
{R. H. Fowler and L. Nordheim}, Proc. R. Soc. London, 
Ser. A {\bf 119}, 173 (1928).

\bibitem{gam28}
{G. Gamov}, Z. Phys. {\bf 51},  204  (1928).

\bibitem{korshe78}
{I. Ya. Korenblit and E. F. Shender}, 
Zh. Eksp. Teor. Fiz. {\bf 75}, 1862 (1978)
[JETP {\bf 48}, 937 (1978)].

\bibitem{chu79}
{E. M. Chudnovsky}, Zh. Eksp. Teor. Fiz. {\bf 77},  2157  (1979)
[JETP {\bf 50}, 1035 (1979)].

\bibitem{enzsch86}
{M. Enz and R. Schilling}, J. Phys. C {\bf 19},  L711  (1986).

\bibitem{chugun88}
{E. M. Chudnovsky and L. Gunther}, Phys. Rev. Lett. {\bf 60},  661  (1988).

\bibitem{lvhsuto86}
{J. L. van Hemmen and A. S\"ut\H o}, Physica B {\bf 141},  37  (1986).

\bibitem{kra40}
{H. A. Kramers}, Physica {\bf 7},  284  (1940).

\bibitem{stowoh4891}
{E. C. Stoner and E. P. Wohlfart}, Philos. Trans. R. Soc. London, Ser. A {\bf 240},  599  (1948); 
reprinted in: IEEE Trans. Magn. {\bf MAG-27},  3475 (1991).

\bibitem{nee49}
{L. N\'{e}el}, Ann. Geophys. (C.N.R.S.) {\bf 5},  99  (1949).

\bibitem{bro63}
{W. F. Brown, Jr.}, Phys. Rev. {\bf 130},  1677  (1963).

\bibitem{haetalbor90}
{P. H\"anggi, P. Talkner, and M. Borkovec}, 
Rev. Mod. Phys. {\bf 62}, 251 (1990).

\bibitem{stachubar92}
{P. C. E. Stamp, E. M. Chudnovsky, and B. Barbara}, 
Int. J. Mod. Phys. B {\bf 6},  1355  (1992).

\bibitem{cofcalwal96book}
{W. T. Coffey, Yu. P. Kalmykov, and J. T. Waldron}, 
{\em {T}he {L}angevin {E}quation} (Word Scientific, Singapore, 1996).

\bibitem{awsetal92}
{D. D. Awshalom, J. F. Smyth, G. Grinstein, D. P. DiVincenzo, and D. Loss}, Phys. Rev. Lett. {\bf 68},  3092  (1992).

\bibitem{gidetal95}
{S. Gider, D. D. Awshalom, T. Douglas, S. Mann, and M. Chaprala}, Science {\bf 268},  77  (1995).

\bibitem{tejetal97}
{J. Tejada, X. X. Zhang, E. del Barco, J. M. Hern\'andez, and E. M.
  Chudnovsky}, Phys. Rev. Lett. {\bf 79},  1754  (1997).

\bibitem{weretal97q}
{W. Wernsdorfer, E. Bonet Orozco, K. Hasselbach, A. Benoit, D. Mailly, O. Kubo, H. Nakano, and B. Barbara}, Phys. Rev. Lett. {\bf 79},  4014  (1997).

\bibitem{novses95}
{M. A. Novak and R. Sessoli},  in {\em Quantum Tunneling of Magnetization}, edited by L. Gunther and B. Barbara (Kluwer, Dordrecht, 1995).

\bibitem{baretal95}
{B. Barbara {\em et. al.}}, J. Magn. Magn. Mater. {\bf 140-144},  1825  (1995).

\bibitem{frisartejzio96}
{J. R. Friedman, M. P. Sarachik, J. Tejada, and R. Ziolo}, 
Phys. Rev. Lett. {\bf 76},  3830  (1996).

\bibitem{heretal96}
{J. M. Hern\'andez, X. X. Zhang, F. Luis, J. Bartolom\'e, J. Tejada, and R. Ziolo}, Europhys. Lett. {\bf 35},  301  (1996).

\bibitem{thoetal96}
{L. Thomas, F. Lionti, R. Ballou, D. Gatteschi, R. Sessoli, and B. Barbara},
Nature (London) {\bf 383},  145  (1996).

\bibitem{popular}
{P. C. E. Stamp}, Nature (London) {\bf 383},  125  (1996);
{E. M. Chudnovsky}, Science {\bf 274},  938  (1996);
{B. Schwarzschild}, Phys. Today {\bf 50}, No. 1, 17  (1997).

\bibitem{sanetal97}
{C. Sangregorio, T. Ohm, C. Paulsen, R. Sessoli, and D. Gatteschi}, 
Phys. Rev. Lett. {\bf 78},  4645  (1997).

\bibitem{gol59}
{V. I. Goldanskii}, Dokl. Akad. Nauk. SSSR {\bf 124},  1261  (1959)
[Sov. Phys. Dokl., {\bf 4}, 74 (1959)].

\bibitem{bel5980}
{R. P. Bell}, Trans. Faraday Soc. {\bf 55},  1  (1959);
{\em The {T}unnel {E}ffect in {C}hemistry} 
(Chapman and Hall, London, 1980).

\bibitem{aff81}
{I. Affleck}, Phys. Rev. Lett. {\bf 46},  388  (1981).

\bibitem{calleg83}
{A. O. Caldeira and A. J. Leggett}, Ann. Phys. (N.Y.) {\bf 149},  374  (1983).

\bibitem{grawei84}
{H. Grabert and U. Weiss}, Phys. Rev. Lett. {\bf 53},  1787  (1984).

\bibitem{graolswei85}
{H. Grabert, P. Olschowski, and U. Weiss}, 
Phys. Rev. B {\bf 32}, 3348 (1985).

\bibitem{larovc84}
{A. I. Larkin and Yu. N. Ovchinnikov}, 
Zh. Eksp. Teor. Fiz. {\bf 86}, 719 (1984)
[JETP {\bf 59}, 420 (1984)].

\bibitem{zwe85}
{W. Zwerger}, Phys. Rev. A {\bf 31},  1745  (1985).

\bibitem{rishaefre85}
{P. S. Riseborough, P. H\"anggi, and E. Freidkin}, 
Phys. Rev. A {\bf 32}, 489 (1985).

\bibitem{larovc83}
{A. I. Larkin and Yu. N. Ovchinnikov}, 
Pis'ma Zh. Eksp. Teor. Fiz. {\bf 37}, 322 (1983)
[JETP Lett. {\bf 37}, 382 (1983)].

\bibitem{chu92}
{E. M. Chudnovsky}, Phys. Rev. A {\bf 46},  8011  (1992).

\bibitem{chugar97}
{E. M. Chudnovsky and D. A. Garanin}, Phys. Rev. Lett.  
{\bf 79}, 4469 (1997).

\bibitem{lan37}
{L. D. Landau}, Z. Phys. Sowjet. {\bf 11},  26  (1937).

\bibitem{morivlbla94}
{C. Morais-Smith, B. Ivlev, and G. Blatter}, 
Phys. Rev. B {\bf 49},  4033  (1994).

\bibitem{gar94}
{J. Garriga}, Phys. Rev. D {\bf 49},  5497  (1994).

\bibitem{fer95}
{A. Ferrera}, Phys. Rev. D {\bf 52},  6717  (1995).

\bibitem{habmottin96}
{S. Habib, E. Mottola, and P. Tinyakov}, Phys. Rev. D {\bf 54},  7774  (1996).

\bibitem{zimtchmue97}
{F. Zimmerschied, D. H. Tchrakian, and H. J. W. M\"uller-Kirsten},
 Forschritte der Physik {\bf 46} 147 (1998).

\bibitem{skv97}
{M. A. Skvortsov}, Phys. Rev. B {\bf 55},  515  (1997).

\bibitem{gorbla97}
{D. A. Gorokhov and G. Blatter}, Phys. Rev. B {\bf 56},  3130  (1997).

\bibitem{scharf74}
{G. Scharf}, Ann. Phys. (N.Y.) {\bf 83},  71  (1974).

\bibitem{zasulytsu83}
{O. B. Zaslavskii, V. V. Ulyanov, and V. M. Tsukernik}, 
Fiz. Nizk. Temp. {\bf 9},  511  (1983)
[Sov. J. Low Temp. Phys. {\bf 9}, 259 (1983)].

\bibitem{schwrelvh87}
{G. Scharf, W. F. Wreszinski, and J. L. van Hemmen}, 
J. Phys. A: Math. Gen. {\bf 20},  4309  (1987).

\bibitem{zas90pla}
{O. B. Zaslavskii}, Phys. Lett. A {\bf 145},  471  (1990).

\bibitem{zas90prb}
{O. B. Zaslavskii}, Phys. Rev. B {\bf 42},  992  (1990).

\bibitem{garchu97}
{D. A. Garanin and E. M. Chudnovsky}, Phys. Rev. B {\bf 56},  11102  (1997).

\bibitem{vilharsesret94}
{J. Villain, F. Hartmann-Boutron, R. Sessoli, and A. Rettori}, Europhys. Lett. {\bf 27},  159  (1994).

\bibitem{gar97pre}
{D. A. Garanin}, Phys. Rev. E {\bf 55},  2569  (1997).

\bibitem{lanlif35}
{L. D. Landau and E. M. Lifshitz}, Z. Phys. Sowjet. {\bf 8},  153  (1935).

\bibitem{gil55}
{T. L. Gilbert}, unpublished report; mentined in: Phys. Rev. {\bf 100},  1243  (1955).

\bibitem{smiroz76}
{D. A. Smith and F. A. de Rozario}, 
J. Magn. Magn. Mater. {\bf 3},  219  (1976).

\bibitem{bro79}
{W. F. Brown, Jr.}, IEEE Trans. Magn. {\bf MAG-15},  1196  (1979).

\bibitem{kligun90}
{I. Klick and L. Gunther}, J. Stat. Phys. {\bf 60},  473  (1990).

\bibitem{geocofmul97}
{L. J. Geoghegan, W. T. Coffey, and B. Mulligan}, 
Adv. Chem. Phys. {\bf 100}, 475  (1997).

\bibitem{cofetal97}
{W. T. Coffey, D. S. F. Crothers, J. L. Dormann, L. J. Geoghegan, and
E. C. Kennedy}, J. Magn. Magn. Mater.  {\bf 173} L219 (1997).

\bibitem{weretal97}
{W. Wernsdorfer, E. Bonet Orozco, K. Hasselbach, A. Benoit, B. Barbara, N. Demoncy, A. Loiseau, and D. Mailly}, 
Phys. Rev. Lett. {\bf 78},  1791 (1997).

\bibitem{raz80}
{M. Razavy}, Am. J. Phys. {\bf 48},  285  (1980).

\bibitem{lanlif3}
{L. D. Landau and E. M. Lifshitz}, {\em Quantum {M}echanics} (Pergamon, London, 1965).

\bibitem{kem35}
{E. C. Kemble}, Phys. Rev. {\bf 48},  549  (1935).

\bibitem{weihae83}
{U. Weiss and W. Haeffner}, Phys. Rev. D {\bf 27},  2916  (1983).

\bibitem{she83}
{H. K. Shepard}, Phys. Rev. D {\bf 27},  1288  (1983).

\bibitem{garg95diss}
{A. Garg}, Phys. Rev. B {\bf 51},  15161  (1995).

\bibitem{garkim8991}
{A. Garg and G.-H. Kim}, Phys. Rev. Lett. {\bf 63},  2512  (1989);
Phys. Rev. B {\bf 43},  712  (1991).

\bibitem{roshanluk95}
{R. Rose, S. Han, and J. E. Lukens}, Phys. Rev. Lett. {\bf 75},  1614  (1995).

\bibitem{silpalrugrus97}
{P. Silvestrini, V. G. Palmieri, B. Ruggiero, and M. Russo}, Phys. Rev. Lett. {\bf 79},  3046  (1997).

\bibitem{aha69}
{A. Aharoni}, Phys. Rev. {\bf 177},  793  (1969).

\bibitem{byrfri79}
{P. F. Byrd and M. D. Friedman}, 
{\em Handbook of {E}lliptic {I}ntegrals for {E}ngineers and {S}cientists} (Springer, New York, 1979).

\bibitem{gar91jpa}
{D. A. Garanin}, J. Phys. A: Math. Gen. {\bf 24},  L61  (1991).

\bibitem{harbou95}
{F. Hartmann-Boutron}, J. Phys. I {\bf 5},  1281  (1995).

\bibitem{fey72}
{R. P. Feynman}, {\em Statistical {M}echanics} (Benjamin, New York, 1972).

\bibitem{migchu96}
{M.-Carmen Miguel and E. M. Chudnovsky}, Phys. Rev. B {\bf 54},  388  (1996).

\bibitem{chufri}
{E. M. Chudnovsky and J. R. Friedman},   (unpublished).

\end{thebibliography}
\end{document}